\documentclass[fleqn,usenatbib]{mnras}

\usepackage{newtxtext,newtxmath}
\usepackage[T1]{fontenc}
\usepackage{ae,aecompl}

\usepackage{graphicx}	% Including figure files
\usepackage{amsmath}	% Advanced maths commands
\usepackage{multicol} 
\usepackage{multirow}
\usepackage{comment}
\usepackage{array}
\usepackage{soul}
\hypersetup{colorlinks=true,citecolor=blue,linkcolor=purple,filecolor=black,runcolor=black,breaklinks=true}
\usepackage{etoolbox}
\usepackage{ae,aecompl}
\usepackage[usenames]{color}
\usepackage{hyperref}
\usepackage{threeparttable}
\usepackage{mathrsfs}
\usepackage{scalerel}
\usepackage{longtable,booktabs,threeparttablex}
\definecolor{purple}{RGB}{160,32,240}

\definecolor{red}{RGB}{225,50,50}

\definecolor{addchange}{RGB}{215,25,25}

\definecolor{removechange}{RGB}{25,25,215}

\newcommand{\HST}{\emph{HST}}

\newcommand{\Spitzer}{\emph{Spitzer}}
\newcommand{\Muv}{\ensuremath{\mathrm{M}_{\mathrm{UV}}^{ }}}

\newcommand{\Lya}{\ensuremath{\mathrm{Ly}\alpha}}
\newcommand{\zLya}{\ensuremath{\mathrm{z}_{_{\mathrm{Ly\alpha}}}}}

\newcommand{\Msol}{\ensuremath{\mathrm{M}_{\odot}}}
\newcommand{\Mstar}{\ensuremath{\mathrm{M}_{\ast}}}
\newcommand{\logMstar}{\ensuremath{\log\left(\mathrm{M}_{\ast}/\mathrm{M}_{\odot}\right)}}
\newcommand{\logLIR}{\ensuremath{\log\left(\mathrm{L}_{\mathrm{IR}}/\mathrm{L}_{\odot}\right)}}

\newcolumntype{P}[1]{>{\centering\arraybackslash}p{#1}}
\newcommand\Tstrut{\rule{0pt}{2.6ex}}         % = `top' strut
\newcommand\Bstrut{\rule[-1.2ex]{0pt}{0pt}}   % = `bottom' strut

\title[Radio$+$Far-IR Emission from a Massive z$\simeq$6.8 Galaxy?]{Radio and Far-IR Emission Associated with a Massive Star-forming Galaxy Candidate at z$\simeq$6.8: A Radio-Loud AGN in the Reionization Era?}

\author[R. Endsley et al.]{
Ryan Endsley$^{1}$\thanks{E-mail: rendsley@email.arizona.edu},
Daniel P. Stark$^{1}$, 
Xiaohui Fan$^{1}$, 
Renske Smit$^{2}$,
Feige Wang$^{1}$,
\newauthor{
Jinyi Yang$^{1,3}$,
Kevin Hainline$^{1}$,
Jianwei Lyu$^{1}$,
Rychard Bouwens$^{4}$,
Sander Schouws$^{4}$}
\\
$^{1}$Steward Observatory, University of Arizona, 933 N Cherry Ave, Tucson, AZ 85721 USA\\
$^{2}$Astrophysics Research Institute, Liverpool John Moores University, 146 Brownlow Hill, Liverpool L3 5RF, United Kingdom\\
$^{3}$Strittmatter Fellow\\
$^{4}$Leiden Observatory, Leiden University, NL-2300 RA Leiden, Netherlands\\
}

\date{Accepted XXX. Received YYY; in original form ZZZ}

\pubyear{2021}

\begin{document}
\label{firstpage}
\pagerange{\pageref{firstpage}--\pageref{lastpage}}
\maketitle

% Abstract of the paper

\begin{abstract}
We report the identification of radio (0.144--3 GHz) and mid-, far-infrared, and sub-mm (24--850$\mu$m) emission at the position of one of 41 UV-bright ($\mathrm{M_{\mathrm{UV}}}^{ }\lesssim-21.25$) $z\simeq6.6-6.9$ Lyman-break galaxy candidates in the 1.5 deg$^2$ COSMOS field.
This source, COS-87259, exhibits a sharp flux discontinuity (factor $>$3) between two narrow/intermediate bands at 9450 \AA{} and 9700 \AA{} and is undetected in all nine bands blueward of 9600 \AA{}, as expected from a Lyman-alpha break at $z\simeq6.8$.
The full multi-wavelength (X-ray through radio) data of COS-87529 can be self-consistently explained by a very massive (M$_{\ast}=10^{10.8}$ M$_{\odot}$) and extremely red (rest-UV slope $\beta = -0.59$) $z\simeq6.8$ galaxy with hyperluminous infrared emission ($\mathrm{L_{IR}}=10^{13.6}$ L$_{\odot}$) powered by both an intense burst of highly-obscured star formation (SFR$\approx$1800 M$_{\odot}$ yr$^{-1}$) and an obscured ($\tau_{_{\mathrm{9.7\mu m}}} = 7.7\pm2.5$) radio-loud (L$_{\mathrm{1.4\ GHz}} \approx 10^{25.4}$ W Hz$^{-1}$) AGN.
The radio emission is compact (1.04$\pm$0.12 arcsec) and exhibits an ultra-steep spectrum between 1.32--3 GHz ($\alpha=-1.57^{+0.22}_{-0.21}$) that flattens at lower frequencies ($\alpha = -0.86^{+0.22}_{-0.16}$ between 0.144--1.32 GHz), consistent with known $z>4$ radio galaxies.
We also demonstrate that COS-87259 may reside in a significant (11$\times$) galaxy overdensity, as common for systems hosting radio-loud AGN.
While we find that low-redshift solutions to the optical+near-infrared data are not preferred, a spectroscopic redshift will ultimately be required to establish the true nature of COS-87259 beyond any doubt.
If confirmed to lie at $z\simeq6.8$, the properties of COS-87259 would be consistent with a picture wherein AGN and highly-obscured star formation activity are fairly common among very massive (M$_{\ast}>10^{10}$ M$_{\odot}$) reionization-era galaxies.
\end{abstract}
% Select between one and six entries from the list of approved keywords.
% Don't make up new ones.

%%%%%%%%%%%%%%%%%%%%%%%%%%%%%%%%%%%%%%%%%%%%%%%%%%

%%%%%%%%%%%%%%%%% BODY OF PAPER %%%%%%%%%%%%%%%%%%

\defcitealias{Endsley2021_OIII}{E21a}

\section{Introduction} \label{sec:intro}

Deep optical and near-infrared imaging surveys conducted with the Hubble Space Telescope (\HST{}) have uncovered over one thousand Lyman-break galaxies at $z>6$ \citep[e.g.][]{McLure2013,Atek2015a,Bouwens2015_LF,Finkelstein2015_LF,Ishigaki2018}, providing our first census of unobscured star formation in the reionization era. 
The galaxies in these deep but small-area surveys ($\sim$0.2 deg$^2$ total) tend to have low UV luminosities (\Muv{} $>$ $-$20) and correspondingly low stellar masses (\Mstar{} $<$ 10$^9$ \Msol{}; e.g. \citealt{Bhatawdekar2019,Stefanon2021_SMF}). 
The rest-UV continuum slopes of these galaxies are blue ($\beta < -2$; e.g. \citealt{Bouwens2014_beta}), consistent with minimal reddening from dust.  
The {\it Spitzer}/IRAC photometry shows evidence for very intense rest-optical emission lines \citep{Labbe2013,Smit2014,Smit2015,deBarros2019,Endsley2021_OIII}, as expected for systems with large specific star formation rates and rapidly rising star formation histories. 
These low luminosity star-forming systems are thought to play a significant role in driving the reionization process (e.g. \citealt{Bouwens2015_reionization,Robertson2015,Ishigaki2018,Finkelstein2019,Naidu2020}). 

Over the last five years, attention has begun to turn to $z>6$ galaxies identified over much wider areas ($>$ 5 deg$^2$) in ground-based imaging datasets (e.g. \citealt{McCracken2012,Jarvis2013,Aihara2019}).  
The wide areas offer several advantages with respect to earlier studies. 
By probing larger volumes, they enable the identification of rare UV-bright (\Muv{} $<$ $-$21.5) galaxies, providing a first census of the most massive  (\Mstar{} $>$ 10$^{10}$ M$_{\odot}$) UV-luminous star-forming systems in the reionization era  \citep[e.g.][]{Bowler2014,Bowler2020,Stefanon2017_Brightestz89,Stefanon2019,Ono2018,Endsley2021_OIII}.  
Because of the brightness of this population ($J$=24--25), they provide an ideal sample for detailed investigation of the gas, dust, and stellar populations in early UV-luminous galaxies. 
Many of these galaxies are likely to trace overdense regions, some of which may have carved out large ionized regions in the mostly neutral IGM. 
The wide-area imaging surveys are large enough to characterize these environments and (via Ly$\alpha$ emission follow-up) begin mapping the likely size of ionized bubbles in their vicinity \citep[e.g.][]{Castellano2018,Endsley2021_LyA,Endsley2022_bubble,Hu2021}. 

Large samples of UV-luminous reionization-era systems have now been identified via standard dropout selection techniques over these wide-area ground-based imaging fields. 
Initial efforts have focused on characterizing the luminosity function \citep{Bowler2014,Bowler2020,Stefanon2017_Brightestz89,Ono2018} and stellar mass function \citep{Stefanon2019} of the UV-bright galaxies.
The broadband SEDs have revealed comparably intense [OIII]+H$\beta$ emission and specific star formation rates as are seen in less luminous galaxies \citep{Endsley2021_OIII}, suggesting broadly similar recent star formation histories.  
The rest-UV colors of the most UV-luminous galaxies tend to be fairly blue (average $\beta = -2$) suggesting typically low dust reddening at the bright end of the UV luminosity function. 
Work is now extending these studies to other wavelengths, building a more complete picture of the gas and dust in early UV-luminous galaxies. 
Most of this progress has come from ALMA, which has proven very effective at detecting far-IR cooling lines (i.e., [CII], [OIII]) and dust continuum emission in the UV-bright population \citep{Matthee2017_CR7,Matthee2019_resolvedUVandCII,Bowler2018,Carniani2018,Carniani2018_Himiko,Smit2018,Hashimoto2019,Bouwens2021_REBELS,Schouws2021}.

An essential next step is to better understand the characteristics of the most massive (\Mstar{} $>$ 10$^{10}$ \Msol{}) reionization-era galaxies, including those which are substantially reddened by dust.
One notable benefit of explicitly studying massive sources is that they would provide an improved census of highly obscured star formation activity in the very early Universe \citep{Casey2018}.
Typical UV-bright galaxies at $z\sim7-8$ often show relatively weak dust continuum emission consistent with obscured SFRs $\lesssim$30 \Msol{} yr$^{-1}$ \citep[e.g.][]{Bowler2018,Bouwens2021_REBELS,Schouws2021}, far lower than that of $z>6$ sub-mm selected systems ($\sim$500--3000 \Msol{} yr$^{-1}$; \citealt{Riechers2013,Marrone2018,Zavala2018}).
The parameter space between these two populations is conceivably occupied by very massive galaxies with exceptionally red rest-UV slopes ($\beta \gtrsim -1$) that can be difficult to identify with broad-band Lyman-break selection techniques.

Detailed studies of very massive reionization-era galaxies may additionally deliver important clues into the properties of early supermassive black holes (SMBHs).
Much attention has been dedicated to answering how the $>$10$^9$ \Msol{} black holes inferred to power several $z>6$ quasars \citep[e.g.][]{Mortlock2011,Wu2015,Banados2018,Yang2020,Wang2021} were able to form at such early times (see \citealt{Inayoshi2020} for a review).
A key observational stepping stone towards answering this question is to expand our view of $z>6$ AGN activity beyond type 1 quasars to include obscured AGN hosted by the broader massive galaxy population \citep{Vito2019,Onoue2021}.
Additional efforts constraining the frequency of various AGN modes (e.g. radio-loud or super-Eddington accretion) in these massive galaxies can provide insight into not only early SMBH growth mechanisms, but also how SMBHs first co-evolved with their host galaxies \citep[e.g.][]{Kormendy2013,Hardcastle2020,Banados2021}.

In this work, we take a step towards better characterizing the obscured star formation and AGN activity within very massive (\Mstar{} $>$ 10$^{10}$ \Msol{}) reionization-era galaxies utilizing deep X-ray through radio data available over the 1.5 deg$^2$ COSMOS field \citep{Scoville2007}.
The sources explored in this study are drawn from a parent sample of 41 Lyman-break galaxy candidates at $z\simeq6.6-6.9$ \citep{Endsley2021_OIII}.
While all these galaxies are UV-bright ($\Muv{} \lesssim -21.25$), the unique narrow-band dropout selection technique employed by \citet{Endsley2021_OIII} nonetheless enables the identification of exceptionally red ($\beta > -1$) dust-enshrouded massive systems at $z\simeq7$.

We report the detection of 1.4 and 3 GHz radio emission at the position of an extremely massive (\Mstar{} = 10$^{10.8}$ \Msol{}) and red ($\beta = -0.59$) galaxy in this $z\simeq6.6-6.9$ sample (COS-87259). 
We also identify spatially-coincident emission in the mid-IR ({\it Spitzer}/MIPS), far-IR ({\it Herschel}/PACS and {\it Herschel}/SPIRE), and sub-mm (JCMT/SCUBA-2). 
The multiwavelength SED of COS-87259 is fit well by a $z\simeq 6.8$ solution, with the infrared emission powered by highly-obscured star formation (SFR$\approx$1800 \Msol{} yr$^{-1}$) and an obscured radio-loud AGN. 
We discuss the possibility of low-redshift solutions and compare the properties of COS-87259 to the sample of known high-redshift radio galaxies \citep[e.g.][]{Saxena2018,Drouart2020,Yamashita2020} as well as hot dust-obscured galaxies \citep[e.g.][]{Stern2014,Assef2015,Fan2016_hotDOG}.
If this system is confirmed to lie at $z\simeq 6.8$, it would add evidence that AGN may be fairly common in the most massive ($>$ 10$^{10}$ \Msol{}) $z>6$ galaxies, and would further support a picture wherein massive galaxies contribute significantly to the cosmic star formation rate budget during reionization.

Throughout this paper, we quote magnitudes in the AB system \citep{OkeGunn1983}, employ a \citet{Chabrier2003} IMF, adopt a flat $\Lambda$CDM cosmology with parameters $h = 0.7$, $\Omega_\mathrm{M} = 0.3$, and $\Omega_\mathrm{\Lambda} = 0.7$, and report 68\% confidence interval uncertainties unless otherwise specified.

\begin{figure}
\includegraphics{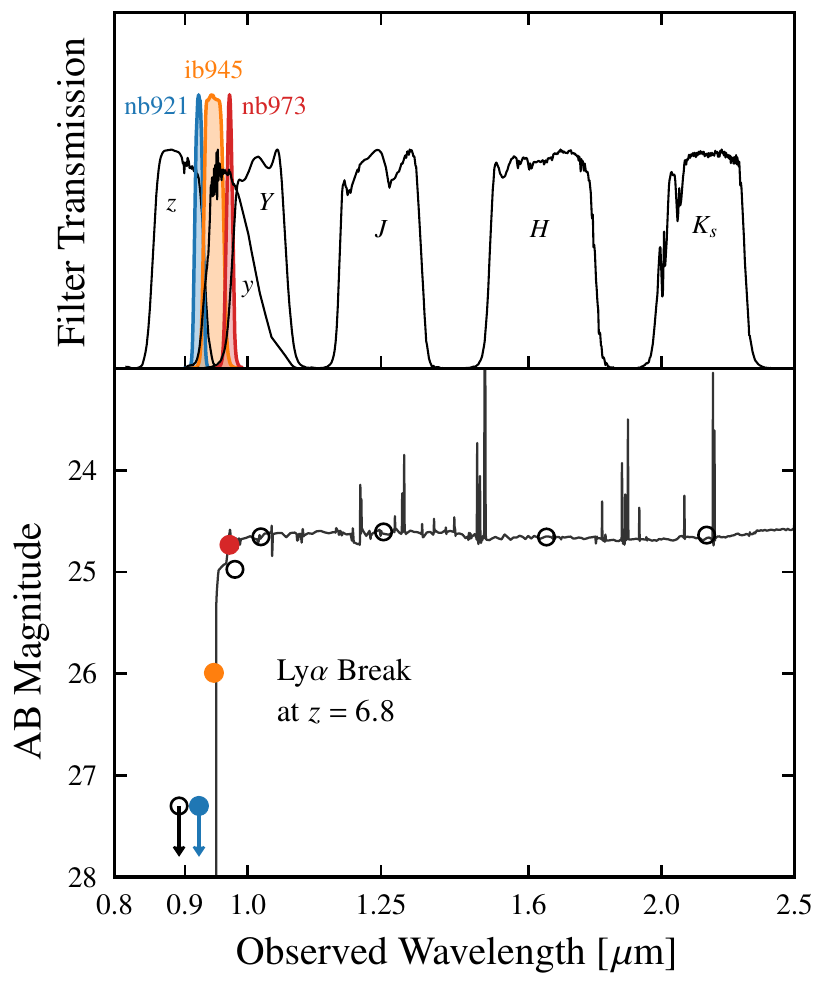}
\caption{Illustration of how narrow/intermediate band data in COSMOS assists in the identification of $z\simeq7$ Lyman-break galaxies. In COSMOS, deep imaging exists in three narrow/intermediate bands covering 0.9--1$\mu$m (HSC nb921, ib945, and nb973). These data effectively act as low-resolution ($R\sim40$) spectra, enabling us to identify sharp flux discontinuities indicative of $z\simeq7$ Ly$\alpha$ breaks. In the top panel, we show the filter transmission curves (arbitrarily normalized) of the bands from Subaru/Hyper Suprime-Cam and VISTA/VIRCam at 0.85--2.5$\mu$m, with the narrow/intermediate band filter curves colored for clarity. In the bottom panel, we show a mock SED (again for illustrative purposes) of a $z=6.8$ galaxy with circles showing the synthetic photometry. A sharp break is evident with the narrow/intermediate-band data.}
\label{fig:filterFigure}
\end{figure}

\section{Lyman-Break Sample Selection} \label{sec:sample}

In this paper, we discuss a sample of $z\simeq 7$ Lyman-break galaxies 
identified over the 1.5 deg$^2$ COSMOS field.
COSMOS has been imaged with Subaru/Hyper Surpime-Cam (HSC; \citealt{Aihara2019,Inoue2020}), VISTA/VIRCam \citep{McCracken2012}, and \Spitzer{}/IRAC \citep{Steinhardt2014,Ashby2018}, providing deep optical and infrared data in 16 filters.
Of particular utility for the selection of reionization-era galaxies is the availability of deep imaging in three narrow/intermediate bands at 0.9--1$\mu$m from HSC (nb921, ib945, and nb973) which cover the wavelength range associated with the Ly$\alpha$ break at $z\simeq7$ (see Fig. \ref{fig:filterFigure}).
With central wavelengths of 9200, 9450, and 9700 \AA{}, these narrow/intermediate-band images yield flux measurements in bins of $\Delta \lambda$ = 250 \AA{} thereby effectively providing low-resolution ($R \sim 40$) spectra in the wavelength range needed for identifying $z\simeq7$ galaxies. 
The improved spectral sampling better pinpoints the observed wavelength where the flux drops off and helps distinguish the sharp flux discontinuities of $z\simeq7$ \Lya{} breaks from the very red, yet relatively smooth spectra of low-redshift dusty galaxies.

Employing this narrow-band dropout technique, \citet[][hereafter \citetalias{Endsley2021_OIII}]{Endsley2021_OIII} identified a sample of $z\simeq6.6-6.9$ UV-luminous (\Muv{} $<$ $-$21) Lyman-break galaxy candidates over COSMOS. This paper presents a possible radio-detected AGN in this sample. To put the initial 
selection of this source in context, we summarize the parent 
sample selection and properties below. For more information, the reader is directed to \citetalias{Endsley2021_OIII}.
The galaxies were selected with the following color cuts\footnote{The HSC ib945 and nb973 data from CHORUS \citep{Inoue2020} were not yet available during the selection analysis of \citetalias{Endsley2021_OIII}. We revisit how these two filters improve the redshift determination of COS-87259 in \S\ref{sec:photoz}.} to identify sharp flux discontinuities expected of Ly$\alpha$ breaks: \textit{z}$-$\textit{y}$>$1.5, \textit{z}$-$\textit{Y}$>$1.5, nb921$-$\textit{Y}$>$1, and \textit{y}$-$\textit{Y}$<$0.4.
As discussed in \citetalias{Endsley2021_OIII}, the nb921 dropout criterion identifies systems with a break redward of 0.92$\mu$m ($z\gtrsim6.6$) while the lack of a strong dropout in \textit{y} limits the position of the \Lya{} break to $\lesssim$0.96$\mu$m, i.e. $z\lesssim6.9$\footnote{The exact redshift interval from any broadband Lyman-break selection is inherently dependent on the assumed \Lya{} equivalent width (EW). Our quoted $z\lesssim$6.9 upper limit assumes \Lya{} EW of 10 \AA{} as typical for UV-luminous $z\simeq7$ galaxies \citep{Endsley2021_LyA}.}. Motivated by sensitivity in IRAC, all galaxies were selected to have apparent magnitudes of \textit{J}$<$25.7 or \textit{K}$_s\!\!<$25.5, corresponding to \Muv{} $\lesssim$ $-$21.25, which is more than 
twice the characteristic UV luminosity at $z\sim7$ \citep{Bowler2017}. 

The \citetalias{Endsley2021_OIII} selection yields a sample of 41 $z\simeq6.6-6.9$ star-forming systems. 
The galaxies are UV-bright with a mean absolute magnitude of M$_{\rm{UV}}=-21.6$ (or 2.5 $\mathrm{L^{\ast}_{UV}}$; \citealt{Bowler2017}). 
The redshift range of the selection places [OIII]+H$\beta$ emission in the IRAC [3.6] bandpass, leaving [4.5] free of strong nebular emission. 
As a result, the degeneracy between nebular emission and old stellar populations is diminished, improving the reliability of the stellar mass inferences. 
\citetalias{Endsley2021_OIII} present derived stellar populations for the sample, revealing an average stellar mass of 10$^9$ M$_\odot$ with values extending up to 2$\times$10$^{10}$ M$_\odot$. 
Spectroscopic follow-up of a subset of these systems have yielded multiple \Lya{} confirmations \citep{Endsley2021_LyA}, demonstrating the efficacy of this narrow-band dropout selection.

\begin{figure}
\includegraphics{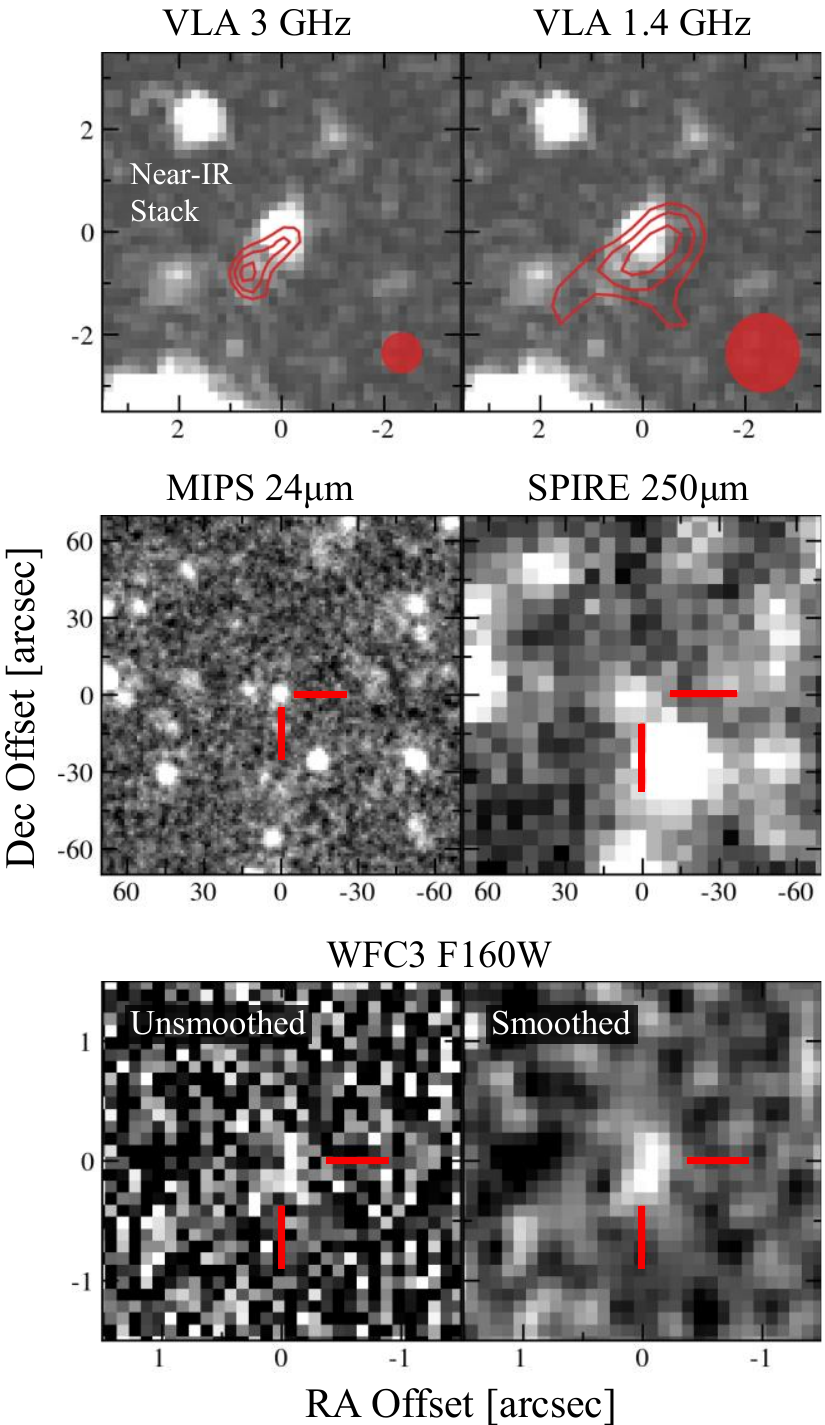}
\caption{Postage stamp cutouts centered at the near-IR position of COS-87259 in the VLA 3 and 1.4 GHz (top), \Spitzer{}/MIPS 24$\mu$m and \textit{Herschel}/SPIRE 250$\mu$m (middle), as well as \HST{} F160W (bottom) bands.
In the top row, the VLA 2, 3, and 4$\sigma$ contours are shown in red overlaid on the near-IR \textit{yYJHK}$_s$ $\chi^2$ stack image. 
The VLA beam size is shown in the bottom right of each panel. 
In the bottom row, we show the unsmoothed WFC3/F160W image on the left while the right image shows the result of smoothing using a 2D Gaussian kernel with $\sigma$=1 pixel (0.1\arcsec{}/pixel). 
The red lines in the middle and bottom rows mark the near-IR centroid of COS-87259 measured from the \textit{yYJHK}$_s$ $\chi^2$ image.}
\label{fig:postageStamps}
\end{figure}

\begin{table}
\centering
\caption{Optical through radio flux density measurements at the position of COS-87259. We report 2$\sigma$ upper limits in cases of non-detections.}
\begin{tabular}{P{1.7cm}P{1.3cm}P{2.3cm}P{1.5cm}} 
\hline
Band & Flux [$\mu$Jy] & Band & Flux [$\mu$Jy] \Tstrut{} \Bstrut{} \\
\hline
HSC \textit{g} & $<$0.017 & IRAC 3.6$\mu$m & 2.75$\pm$0.10 \Tstrut{} \\[3pt]
HSC \textit{r} & $<$0.022 & IRAC 4.5$\mu$m & 2.77$\pm$0.12 \\[3pt]
HSC nb718 & $<$0.056 & IRAC 5.8$\mu$m & 4.42$\pm$1.04 \\[3pt]
HSC \textit{i} & $<$0.027 & IRAC 8.0$\mu$m & $<$4.88 \\[3pt]
ACS F814W & $<$0.027 & MIPS 24$\mu$m & 179.7$\pm$6.4 \\[3pt]
HSC nb816 & $<$0.054 & PACS 100$\mu$m & 5100$\pm$1370 \\[3pt]
HSC \textit{z} & $<$0.030 & PACS 160$\mu$m & 10130$\pm$4260 \\[3pt]
HSC nb921 & $<$0.051 & SPIRE 250$\mu$m & 8830$\pm$1120 \\[3pt]
HSC ib945 & $<$0.062 & SPIRE 350$\mu$m & 8090$\pm$1600 \\[3pt]
HSC nb973 & 0.20$\pm$0.06 & SPIRE 500$\mu$m & 11130$\pm$2220 \\[3pt]
HSC \textit{y} & 0.18$\pm$0.04 & SCUBA-2 850$\mu$m & 6300$\pm$1650 \\[3pt]
VIRCam \textit{Y} & 0.22$\pm$0.04 & VLA 3 GHz & 20.0$\pm$2.6 \\[3pt]
VIRCam \textit{J} & 0.37$\pm$0.05 & VLA 1.4 GHz & 96.3$\pm$14.4 \\[3pt]
WFC3 F160W & 0.44$\pm$0.12 & MeerKAT 1.32 GHz & 71.3$\pm$8.5 \\[3pt]
VIRCam \textit{H} & 0.56$\pm$0.05 &  LOFAR 144 MHz & 475$\pm$180 \\[3pt]
VIRCam \textit{K}$_s$ & 0.67$\pm$0.09 & - & - \\[3pt]
\hline
\end{tabular}
\label{tab:photometry}
\end{table}

\section{Multi-wavelength Observations} \label{sec:observations}

While investigating the physical properties of the most massive galaxies in the \citetalias{Endsley2021_OIII} COSMOS sample, we noted the presence of radio emission at the position of 
COS-87259 (see Fig. \ref{fig:postageStamps}; RA = 09:58:58.27, Dec = $+$01:39:20.2) in the public 1.4 and 3 GHz VLA mosaics \citep{Schinnerer2007,Smolcic2017}. We subsequently found that this source appears to have 
detections in data from \Spitzer{}/MIPS, \textit{Herschel}/PACS, \textit{Herschel}/SPIRE, and JCMT/SCUBA-2 (Fig. \ref{fig:postageStamps}). 
This is the only source in the \citetalias{Endsley2021_OIII} sample that is detected in \Spitzer{}/MIPS, \textit{Herschel}, JCMT, or VLA data.
\citetalias{Endsley2021_OIII} report properties of the source, revealing it to be among the most massive and oldest objects in the photometric sample, with significant UV-optical reddening from dust. 
We will revisit these properties in \S\ref{sec:analysis}. 
Before doing so, we first describe existing observations of this galaxy, detailing the optical/near-infrared (\S\ref{sec:optnear-IR}), radio (\S\ref{sec:radio}), mid/far-infrared (\S\ref{sec:fIR}), and X-ray (\S\ref{sec:Xray}) flux measurements. 
In \S\ref{sec:MMT}, we detail MMT optical spectroscopy of COS-87259. 
Table \ref{tab:photometry} summarizes the multi-wavelength flux density measurements.

\subsection{Optical to Near-Infrared Observations of COS-87259} \label{sec:optnear-IR}

COS-87259 is covered by deep optical and near-infrared (0.3--5$\mu$m) observations from Subaru/HSC, VISTA/VIRCam, and \Spitzer{}/IRAC \citep{McCracken2012,Steinhardt2014,Ashby2018,Aihara2019,Inoue2020}.
We measure the optical/near-IR flux densities using the approach of \citetalias{Endsley2021_OIII} where HSC and VIRCam fluxes are computed in 1.2\arcsec{} diameter apertures while the IRAC 3.6 and 4.5$\mu$m fluxes are computed in 2.8\arcsec{} diameter apertures due to the broader PSF.
When calculating the IRAC photometry, we employ a deconfusion algorithm that first estimates the flux profiles of neighboring sources from the \textit{yYJHK}$_s$ $\chi^2$ prior and then subtracts off those neighboring flux profiles from the IRAC image (see \citetalias{Endsley2021_OIII} for further details).
We find minimal confusion for COS-87259 in the IRAC bands as neighboring sources are inferred to only contribute 3--6\% of the flux within the 2.8\arcsec{} diameter apertures.
All optical and near-IR flux density measurements are aperture corrected using the curve of growth calculated from nearby stars \citepalias{Endsley2021_OIII}.

COS-87259 is undetected in all eight HSC bands blueward of nb973 (\textit{g}, \textit{r}, \textit{i}, \textit{z}, nb718, nb816, nb921, and ib945).
The 2$\sigma$ upper limiting flux densities in these bands range from $<$0.017 $\mu$Jy in \textit{g} to $<$0.062 $\mu$Jy in ib945 (Table \ref{tab:photometry}).
Upon stacking the images of all eight bands we find a total S/N=0.68 in a 1.2\arcsec{} diameter aperture centered on COS-87259, consistent with no flux blueward of $\approx$9450 \AA{}.
COS-87259 is then detected in nb973 with a flux density of 0.20$\pm$0.06 $\mu$Jy which matches that measured in the slightly redder HSC \textit{y} and VIRCam \textit{Y} bands (0.18--0.22 $\mu$Jy).
As discussed further in \S\ref{sec:analysis}, this sharp flux discontinuity between the ib945 and nb973 bands ($>$3$\times$ over 250~\AA) is consistent with expectations of a Ly$\alpha$ break associated with a 
$z\simeq6.8$ galaxy.

The continuum flux density is found to rise smoothly between the \textit{Y} (0.22$\pm$0.04 $\mu$Jy) and \textit{K}$_s$ bands (0.67$\pm$0.09 $\mu$Jy), with a best-fitting spectral slope of F$_{\nu} \propto \lambda^{1.41}$ across the four VIRCam bands.
Extrapolating this slope to redder wavelengths, we would expect to measure flux densities of 1.52 and 2.12 $\mu$Jy in the IRAC 3.6 and 4.5$\mu$m bands, respectively.
We instead measure significantly higher flux densities of 2.75$\pm$0.10 and 2.77$\pm$0.12 $\mu$Jy, respectively.
As we will demonstrate in \S\ref{sec:analysis}, these relatively large IRAC flux densities are consistent with a prominent Balmer break at $z\simeq6.8$, and the VIRCam colors are consistent with a red rest-UV continuum slope. 

COS-87259 has also been observed with \HST{} using the ACS F814W \citep{Scoville2007_HST,Koekemoer2007,Massey2010} and WFC3 F160W \citep{Mowla2019} bands.
We astrometrically correct each archival \HST{} image to the Gaia frame using the IRAF \textsc{ccmap} package.
We find no detection in the F814W image at the position of COS-87259 and measure a 2$\sigma$ upper limiting flux density of $<$0.027 $\mu$Jy in an 0.4\arcsec{} diameter aperture.
This is consistent with the non-detections in HSC \textit{i} and \textit{z} which have similar upper limits.
The F160W image shows a detection (Fig. \ref{fig:postageStamps}) with a flux density of 0.44$\pm$0.12 $\mu$Jy (0.6\arcsec{} aperture), consistent within uncertainties with that measured in VIRCam \textit{H}.
The F160W emission appears slightly extended along the N-S axis with an estimated deconvolved size of 0.41\arcsec{}.

We have visually verified that all HSC, \HST{}, VIRCam, and IRAC data of COS-87259 appear robust.
The local background in all images is smooth and there is no evidence of artificial features or diffraction spikes affecting the photometric measurements.

\begin{figure}
\includegraphics{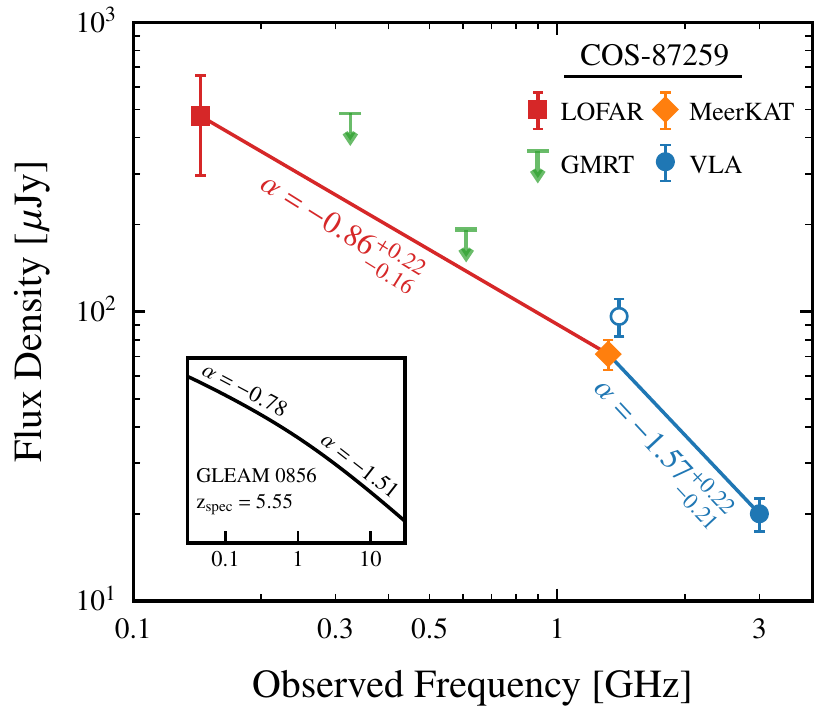}
\caption{Illustration of constraints on the spectral slopes ($S_{\nu} \propto \nu^{\alpha}$) of the radio emission coincident with COS-87259. The MeerKAT 1.32 GHz (orange diamond; \citealt{Heywood2022}) and VLA 3 GHz (blue circle; \citealt{Smolcic2017}) flux densities reveal an ultra-steep spectrum ($\alpha < -1.3$) at high frequencies. The LOFAR 144 MHz flux density measurement (red square; \citealt{Shimwell2022}) indicates that the radio slope of COS-87259 flattens significantly at lower frequencies. This flattening is supported by non-detections in the GMRT 325 and 610 MHz bands (green arrows; \citealt{Tisanic2019}) and is consistent with the observed properties of $z>4$ radio galaxies \citep[e.g.][]{Ker2012,Saxena2018_faintRadio,Saxena2018,Yamashita2020}. As an example, we show the radio SED of GLEAM 0856 at $z=5.55$ in the inset panel which shows an ultra-steep ($\alpha = -1.51$) radio slope at $\gtrsim$1.4 GHz yet a relatively flat slope ($\alpha = -0.78$) at $\lesssim$1 GHz \citep{Drouart2020}. The open blue circle shows the VLA 1.4 GHz flux density measurement \citep{Smolcic2017} which has lower S/N relative to the MeerKAT measurement.}
\label{fig:radioSlope}
\end{figure}

\subsection{Radio Observations of COS-87529} \label{sec:radio}

Deep radio observations at 1.4 and 3 GHz have been conducted over the COSMOS field from the VLA-COSMOS survey \citep{Schinnerer2007,Smolcic2017}. The 1.4 and 3 GHz mosaics have beam sizes of FWHM = 1.5\arcsec{} and 0.75\arcsec{}, respectively, as well as estimated absolute astrometric precisions of 0.055\arcsec{} and 0.01\arcsec{}. 
We identify significant emission near the position of COS-87259 in both the 1.4 and 3 GHz mosaics (Fig. \ref{fig:postageStamps}), where the S/N at the peak of surface brightness profile is 4.8 and 4.5 in the respective images.
Given the significance of these detections, we expect their astrometric precision to be dominated by relative accuracy, which we estimate as (beam FWHM)/(peak S/N).
This results in an astrometric uncertainty of 0.31\arcsec{} and 0.17\arcsec{} for the 1.4 and 3 GHz detections, respectively.
The 3$\sigma$ contours of the VLA detections overlap with the near-infrared centroid of COS-87259 within these 1$\sigma$ uncertainties.
We measure the near-IR centroid from the \textit{yYJHK}$_s$ $\chi^2$ stack (see \citetalias{Endsley2021_OIII}) where the HSC and VIRCam images have been calibrated to the absolute reference frame from Gaia. There are no other sources detected in the near-IR stack that overlap with the 3$\sigma$ VLA contours within the 2$\sigma$ positional uncertainties (Fig. \ref{fig:postageStamps}).

We calculate the VLA flux densities using the \textsc{blobcat} program \citep{Hales2012}, finding $S_{1400} = 96.3 \pm 14.4$ $\mu$Jy and $S_{3000} = 20.0 \pm 2.6$ $\mu$Jy.
These measurements indicate an ultra-steep ($\alpha < -1.3$) radio power spectrum between the 1.4 and 3 GHz bands of $\alpha^{1400}_{3000} = -2.06^{+0.27}_{-0.25}$ where $S_{\nu} \propto \nu^{\alpha}$ (Fig. \ref{fig:radioSlope}).
After fitting the surface brightness profile to a 2D Gaussian, we find that the detection in the 3 GHz image is resolved with an observed length of 1.28$\pm$0.10 arcsec along the major axis, translating to a deconvolved size of 1.04$\pm$0.12 arcsec.
These radio slope and size measurements indicate that this is a compact ($<$20 kpc) steep-spectrum radio source \citep[e.g.][]{ODea1998}, where this $<$20 kpc upper limit holds at any redshift between $z=0-30$.
While the 1.4 GHz data has slightly poorer angular resolution, the detection in this band also appears slightly resolved with an estimated deconvolved size of 1.18$\pm$0.24\arcsec{} along the same major axis as in the 3 GHz image.
We note that the 3 GHz measurement may be slightly underestimating the total flux density given that the source is moderately resolved in this map ($\theta_\mathrm{obs} / \theta_\mathrm{beam} \approx 1.7$)

Very recent data from the South African MeerKAT telescope provides further insight into the radio luminosity of this source. 
Using the Early Science Release products from the MeerKAT International Gigahertz Tiered Extragalactic Explorations (MIGHTEE) survey \citep{Heywood2022}, we find that this source is detected at 1.32 GHz with a flux density of 71.3$\pm$8.5 $\mu$Jy (S/N = 8.4). 
Here we are using the peak flux density measurement given that the radio source is unresolved in the MIGHTEE data (beam FWHM$\approx$9 arcsec).
This higher signal-to-noise 1.32 GHz measurement from MIGHTEE indicates a slightly lower flux density relative to that obtained from the VLA-COSMOS data at 1.4 GHz, resulting in a flatter (though still ultra-steep) slope to high frequencies ($\alpha^{1320}_{3000} = -1.57^{+0.22}_{-0.21}$).

The COSMOS field has also been observed at 144 MHz as part of the LOw-Frequency ARray (LOFAR) Two-metre Sky Survey (LoTSS; \citealt{Shimwell2017,Shimwell2019}).
We use the data from LoTSS DR2 which was processed following the methods described in \citet{Shimwell2022}, resulting in a beam size of FWHM$\approx$6 arcsec.
The LoTSS data reveal a low significance detection (2.6$\sigma$) spatially coincident with COS-87259.
Because this LoTSS detection is unresolved (as expected given the $\approx$1\arcsec{} source size in the VLA maps), we adopt the peak flux density measurement of 475$\pm$180 $\mu$Jy where the error is estimated from the local root-mean-square.
If we extrapolate our measured 1.32--3 GHz spectral slope of $\alpha^{1320}_{3000} = -1.57^{+0.22}_{-0.21}$ to the LOFAR frequency (144 MHz), we would infer a flux density of $S_{144} = 2300^{+1560}_{-1000}$ $\mu$Jy.
This is nearly 5$\times$ larger than that measured from the LOFAR data, indicating that the radio slope of COS-87259 flattens considerably at lower frequencies to $\alpha^{144}_{1320} = -0.86^{+0.22}_{-0.16}$.

Additional data at intermediate frequencies (0.3--0.6 GHz) provide further constraints on the shape of the radio spectra from COS-87259. 
COSMOS has been observed with the GMRT at 325 and 610 MHz where the resulting maps have median 5$\sigma$ sensitivities of 485 $\mu$Jy beam$^{-1}$ and 192 $\mu$Jy beam$^{-1}$, respectively \citep{Tisanic2019}.
No source is present within 10 arcsec of COS-82759 in the published catalog of either GMRT band, suggesting that the radio slope flattens at an observed frequency of $\lesssim$1 GHz.
Assuming the 325 and 610 MHz flux densities are lower\footnote{Given the much larger beam size of the GMRT maps (4--10\arcsec{}), we assume that this low-frequency radio emission will not be resolved. We also note that variability is unlikely impact the differences in radio slope at low vs. high frequencies given that compact steep-spectrum sources are among the least variable types of radio AGN \citep{ODea1998}.} than the median 5$\sigma$ sensitivities, we calculate the following limits: $\alpha^{325}_{1320} > -1.37$ and $\alpha^{610}_{1320} > -1.30$ (Fig. \ref{fig:radioSlope}).
Both of these limits are consistent with the low-frequency slope measurement described above ($\alpha^{144}_{1320} = -0.86^{+0.22}_{-0.16}$).
As we discuss in \S\ref{sec:analysis}, these observed radio properties are consistent with expectations of high-redshift ($z>4$) radio galaxies.

\subsection{Mid-, Far-Infrared, and Sub-mm Observations of COS-87259 } \label{sec:fIR}

The COSMOS field has been observed in the mid-infrared \Spitzer{}/IRAC 5.8 and 8.0$\mu$m bands from the S-COSMOS survey \citep{Sanders2007}.
We take the 5.8 and 8.0$\mu$m photometry of COS-87259 from the S-COSMOS catalog\footnote{We use our own photometric measurements for the 3.6 and 4.5$\mu$m bands because we incorporate more recent and much deeper data in these lower-wavelength IRAC filters (\S\ref{sec:optnear-IR}).} using the 2.9\arcsec{} diameter aperture values and applying the reported\footnote{\url{https://irsa.ipac.caltech.edu/data/COSMOS/gator_docs/scosmos_irac_colDescriptions.html}} aperture corrections.
The S-COSMOS catalog position is separated by only 0.08\arcsec{} from the near-IR position of COS-82759.
A low-significance (4.3$\sigma$) detection is reported in the 5.8$\mu$m band with a flux density of 4.42$\pm$1.04 $\mu$Jy.
COS-87259 is reported to be undetected ($<$2$\sigma$) in the 8.0$\mu$m band with a 2$\sigma$ upper limiting flux of $<$4.88 $\mu$Jy.

Additional deep mid+far-infrared observations have been conducted over the COSMOS field with the \Spitzer{}/MIPS 24$\mu$m, \textit{Herschel}/PACS 100 and 160$\mu$m, and \textit{Herschel}/SPIRE 250, 350, and 500$\mu$m bands \citep{LeFloch2009,Lutz2011,Oliver2012}.
We cross-reference the near-IR position of COS-87259 with the \textit{Herschel} Extragalactic Legacy Project (HELP) catalog containing confusion-corrected flux density measurements in each mid+far-infrared band \citep{Hurley2017,Shirley2019,Shirley2021}.
A source is listed in the HELP catalog 0.12\arcsec{} away from the near-IR centroid of COS-87259 (both the catalog and near-IR images are calibrated to the Gaia 
frame). This source has reported $\geq$5$\sigma$ detections in the MIPS 24$\mu$m band (0.18 mJy) as well as the SPIRE 250, 350, and 500$\mu$m bands (8.1--11.1 mJy).
We  visually identify detections in the 24$\mu$m and 250$\mu$m mosaics (Fig. \ref{fig:postageStamps}) which are less confused than the 350 and 500$\mu$m images.
The HELP catalog also reports lower significance (2.4--3.7$\sigma$) detections in the PACS 100 and 160$\mu$m bands (5.1--10.1 mJy).

We additionally investigated the JCMT/SCUBA-2 850$\mu$m catalog from the SCUBA-2 COSMOS survey (S2COSMOS; \citealt{Simpson2019}).
A 3.8$\sigma$ detection is reported 1.1\arcsec{} away from the near-infrared centroid of COS-87259.
This offset is consistent with the 1$\sigma$ positional accuracy of 3.9\arcsec{}, calculated from the average S2COSMOS beam size (FWHM=14.9\arcsec{}; \citealt{Simpson2019}) and the  detection significance quoted above. 
The 850$\mu$m flux density in the S2COSMOS catalog\footnote{We adopt the total instrumental plus confusion noise reported in the S2COSMOS catalog. However, we use the S2COSMOS flux density measurement that has not been corrected for flux boosting. This is to maintain consistency with the \textit{Herschel} flux densities reported in the HELP catalog which, to the best of our knowledge, have not been corrected for flux boosting and are reported to a similar significance as the SCUBA-2 detection.} is reported as 6.3$\pm$1.65 mJy.

\subsection{X-ray Observations of COS-87259} \label{sec:Xray}

The \textit{Chandra} COSMOS Legacy Survey \citep{Civano2016} has provided deep ($\approx$160 ks) X-ray imaging over the central 1.5 deg$^2$ containing COS-87259.
We checked both the \textit{Chandra} COSMOS Legacy Survey catalog and the \textit{Chandra} Source Catalog (v2.0; \citealt{Evans2019}), finding no source within 1 arcminute of COS-87259.
As an upper limit on the X-ray flux, we adopt the `flux\_sens\_true\_b' value from the \textit{Chandra} Source Catalog (1.8$\times$10$^{-15}$ erg/s/cm$^2$) which is the estimated energy flux required for a point source to be detected and classified as `TRUE' in the 0.5--7.0 keV band at the position of COS-87259. 
We discuss how this X-ray upper limit is consistent with a $z\simeq7$ AGN solution for COS-87259 in \S\ref{sec:self_consistent}.

\subsection{MMT Spectroscopy of COS-87259} \label{sec:MMT}

With the goal of constraining possible \Lya{} emission from COS-87259, we have also followed up this source with the MMT/Binospec spectrograph \citep{Fabricant2019}.
Our observations were conducted in long-slit mode utilizing a slit width of 1.0\arcsec{} and the 600 l/mm grating, resulting in a resolving power of $R\sim4400$.
We adopted a central wavelength of 8750 \AA{} proving spectral coverage betweeen 7490--10010 \AA{} which translates to a \Lya{} ($\lambda_{\mathrm{rest}} = 1215.67$ \AA{}) redshift range of \zLya{} = 5.16--7.23.
During the observations, the longslit was oriented to also contain a star which we use for seeing measurements and flux calibration.

COS-87259 was observed on April 30 and May 4, 2021 under mostly clear conditions but with variable seeing.
To minimize the contribution of exposures with poor seeing, we restrict our analysis to the 3 hours of data (twelve 15 minute exposures) with seeing $<$1.3\arcsec{}.
Our data reduction method largely follows that described in \citet{Endsley2021_LyA} which we briefly describe here.
We reduce each individual exposure using the Binospec data reduction pipeline \citep{Kansky2019} and subsequently co-add the data by applying the weighting scheme of \citet{Kriek2015}.
This weighting approach accounts for differences in seeing and sky transmission between exposures using the amplitude of the 1D flux profile of the slit star as the relative weight of each exposure.
Absolute flux calibration is then performed by determining the factor necessary to match the flux of the star to its PSF \textit{z}-band magnitude taken from the Pan-STARRS catalog \citep{Chambers2016_PanSTARRS}.
We use optimal extraction \citep{Horne1986} to obtain the 1D spectra of both the star and COS-87259 where the seeing from the 3 hours of co-added data is 1.03\arcsec{}.

No emission feature is detected from COS-87259 in our Binospec spectra.
To place an upper limit on the \Lya{} flux, we assume an extraction width equivalent to 220 km s$^{-1}$ (the typical \Lya{} FWHM of $z\simeq7$ galaxies; e.g. \citealt{Pentericci2018,Endsley2021_LyA}).
The resulting 5$\sigma$ \Lya{} flux limit in skyline-free regions of the spectrum is $<$5.3$\times$10$^{-18}$ erg/s/cm$^2$.
Adopting the VIRCam \textit{Y}-band measurement as the continuum flux, this translates to a rest-frame EW upper limit of $<$12.5 \AA{} in clear regions of the spectrum.
As discussed in \S\ref{sec:analysis}, this \Lya{} EW upper limit is consistent with a $z\simeq7$ solution given the typical \Lya{} EWs found for UV-bright $z\simeq7$ galaxies and the extremely dusty nature of this source.

We have also conducted near-infrared spectroscopy with MMT/MMIRS \citep{McLeod2012} to help assess the plausibility of low-redshift galaxy solutions where H$\alpha$ emission may dominate the measured \textit{J}-band photometry (see \S\ref{sec:photoz}).
We used \textsc{mmirsmask} to design a slit mask containing COS-87259 along with several photometrically-selected $z\sim1.5$ extreme emission line galaxies (following \citealt{Tang2019}) with a slit length of 10 arcsec for each object.
Observations were conducted using the \textit{zJ} filter plus the \textit{J} grism resulting in a resolving power of $R=960$ with the adopted 1.0 arcsec slit width, as well as a wavelength coverage of 0.95--1.43$\mu$m for COS-87259. 
We obtained 1.33 hours (16$\times$300 s) of exposure time under clear conditions with an average 0.76 arcsec seeing on December 17, 2021. 
Because conditions were stable throughout the observations, we use the non-weighted stack of the spectra automatically produced using the MMIRS reduction pipeline \citep{Chilingarian2015} provided by the SAO Telescope Data Center.
We perform optimal 1D extraction \citep{Horne1986} and use UltraVISTA \textit{J}-band photometric measurements of nearby slit stars for the absolute flux calibration.

We do not identify any emission feature from COS-87259 in our MMIRS data.
Notably, the observations did yield several line detections for photometrically-selected $z\sim1.5$ extreme emission line galaxies (Tang et al. in prep).
In skyline-free regions of the spectrum, the 5$\sigma$ limiting flux is 8.0$\times$10$^{-18}$ erg/s/cm$^2$ adopting an extraction width of 200 km/s.
As discussed in \S\ref{sec:photoz}, the combination of non-detections in both our Binospec and MMIRS data suggest that COS-87259 is unlikely to be a dusty, low-redshift extreme emission line galaxy.

\begin{figure}
\includegraphics{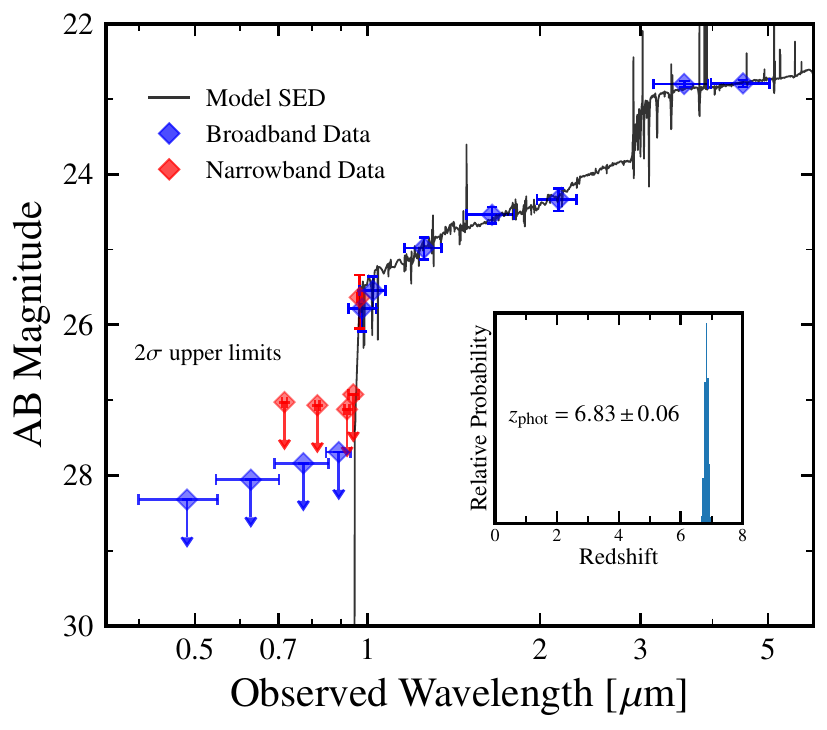}
\caption{Results from fitting the optical and near-IR (0.3--5$\mu$m) photometry of COS-87259 with \textsc{beagle} \citep{Chevallard2016}. Broad-band measurements are shown with blue diamonds while the narrow/intermediate-band data from HSC are shown in red. The best-fitting model SED is shown with the black curve. The redshift of this source is highly favored at $z>6$ ($>$99.99\% confidence; see inset panel) due to the sharp flux discontinuity between the HSC ib945 and nb973 bands. Because the Lyman-alpha break is expected to lie in the small wavelength range between these two bands ($\Delta \lambda = 250$ \AA{}), the redshift is precisely constrained to $z=6.83\pm0.06$. We compare the fits of forced low-redshift solutions in Fig. \ref{fig:compareLowz}.}
\label{fig:BEAGLE}
\end{figure}

\begin{figure*}
\includegraphics{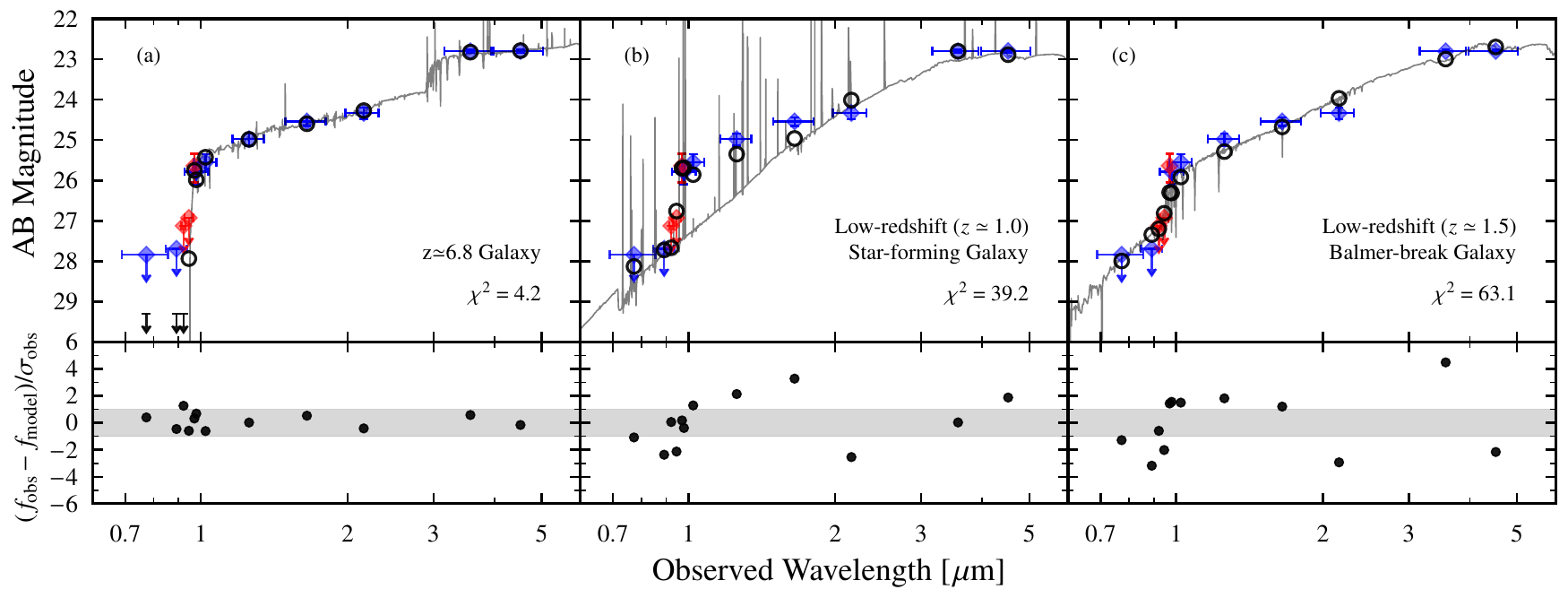}
\caption{Comparison of (a) the $z\simeq6.8$ best-fitting solution from \textsc{beagle} versus that of (b) a low-redshift star-forming galaxy, and (c) a low-redshift galaxy with an old stellar population yielding a prominent Balmer break. In the top panels, the observed data is shown with colored markers similar to Fig. \ref{fig:BEAGLE} while the synthetic photometry from the models are shown with open circles. Only a $z\simeq6.8$ galaxy model is able to well reproduce the sharp flux discontinuity between the ib945 and nb973 bands as well as the observed near-infrared data. In the lower panels, we show how well the synthetic photometry from each best-fitting model matches the observed data. Both low-redshift models yield multiple 2--5$\sigma$ offsets. The lack of emission line detections in our MMT/Binospec spectrum adds independent evidence that the low-redshift star-forming galaxy solution is unlikely (see text). For clarity, we only show a subset of the available optical/near-IR bands but the reported $\chi^2$ values incorporate all HSC, VISTA, and IRAC data.}
\label{fig:compareLowz}
\end{figure*}

\section{The Physical Properties of COS-87529} \label{sec:analysis}

In this section, we discuss the possible origin of the optical through radio emission seen from COS-87259. 
We first use the new HSC ib945 and nb973 data to revisit the optical/near-IR photometric redshift of this system and discuss whether the observed mid-IR, far-IR, and radio emission may be originating from a low-redshift source (\S\ref{sec:photoz}).
We then explore whether the full X-ray through radio data for COS-87259 can be self-consistently explained by a $z\simeq7$ solution (\S\ref{sec:self_consistent}).
Finally, we investigate whether COS-87259 may trace a galaxy overdensity at $z\simeq7$, as is common for radio galaxies at lower redshifts (\S\ref{sec:overdensity}).

\subsection{Photometric Redshift of COS-87529 and Possibility of Low-redshift Interpretation} \label{sec:photoz}

COS-87259 was originally selected as a $z\simeq6.6-6.9$ Lyman-break galaxy candidate utilizing photometric measurements in four filters covering 0.9--1$\mu$m (HSC \textit{z}, nb921, \textit{y}, and VIRCam \textit{Y}; see \S\ref{sec:sample}).
Since the original identification of this source, additional deep HSC ib945 and nb973 imaging has become available over COSMOS, providing further information on the strength and wavelength position of the spectral break in this source.
Here, we revisit the photometric redshift of COS-87259 using this new narrow/intermediate-band data and discuss the possibility of a  low-redshift solution. 

We quantify the photometric redshift of COS-87259 by fitting its optical/near-IR (0.3--5$\mu$m) photometry with the BayEsian Analysis of GaLaxy sEds (\textsc{beagle}; \citealt{Chevallard2016}) SED fitting code.
\textsc{beagle} adopts the \citet{Gutkin2016} photoionization models of star-forming galaxies and calculates the posterior probability distribution of galaxy properties by employing the Bayesian \textsc{multinest} algorithm \citep{Feroz2008,Feroz2009}.
Our fitting procedure with \textsc{beagle} is identical to that described in \citetalias{Endsley2021_OIII}, with the exception that we now include the HSC ib945 and nb973 flux density measurements.
Briefly, we allow the photometric redshift to vary between $0<z<10$ where absorption from intergalactic HI is applied using the empirical model of \citet{Inoue2014}.
We adopt a \citet{Chabrier2003} IMF and model dust attenuation using the SMC extinction curve of \citet{Pei1992}.
We have verified that our conclusions do not change significantly if we instead adopt the \citet{Calzetti2000} dust law.
The minimum allowed age of the stellar population is 1 Myr where we assume a delayed star formation history (SFR $\propto t\, e^{-t/\tau}$) and allow for a recent ($<$10 Myr) burst of star formation with specific star formation rate between 0.1--1000 Gyr$^{-1}$.
We refer the interested reader to \citetalias{Endsley2021_OIII} for further details of our SED fitting procedure.

With \textsc{beagle}, we find that the combined optical/near-IR data are well described by a galaxy at $z = 6.83 \pm 0.06$ (Fig. \ref{fig:BEAGLE}).
At this redshift, the factor of $>$3 increase in flux density between the HSC ib945 and nb973 bands ($\Delta \lambda$ = 250 \AA{}) is expected from a Lyman-alpha break caused by the IGM (e.g. \citealt{Madau1995,Inoue2014}).
The observed red VIRCam colors yield a measured rest-UV slope of $\beta = -0.59$ (if at $z\simeq6.8$), suggesting strong dust attenuation with $\approx$2 magnitudes of extinction at rest-frame 1500 \AA{}.
In the best-fitting \textsc{beagle} model (Fig. \ref{fig:BEAGLE}), the flux excess in the IRAC bands is explained by the presence of a relatively old ($\sim$300 Myr) stellar population producing a strong Balmer break.
However, there are comparably well-fit $z\simeq6.8$ models where the rest-optical emission is dominated by light from a young ($\lesssim$10 Myr) stellar population that is more heavily obscured by dust.
We come back to discuss the nature of COS-87259 (assuming a $z\simeq6.8$ solution) in \S\ref{sec:self_consistent} after incorporating the longer-wavelength (mid-IR, far-IR, and radio) data which better anchor the dust-obscured star formation rate and stellar mass of this system.

It is important to additionally explore whether the data can be adequately explained by a low-redshift ($z<4$) solution. 
We first investigate whether the optical/near-IR data are consistent with emission from a single low-redshift galaxy by re-running \textsc{beagle} forcing the redshift to $z<4$. 
Here, we allow for a larger variety of dust extinction prescriptions: SMC \citep{Pei1992}, \citet{Calzetti2000}, \citet{CharlotFall2000}, and \citet{Chevallard2013}.
Two of the most likely low-redshift galaxy contaminants of Lyman-break selections are dusty, young galaxies with strong emission lines and galaxies with old stellar populations which produce a prominent Balmer break.
Neither of these low-redshift solutions are able to reproduce the observed optical/near-IR photometry in the context of the \textsc{beagle} models (Fig. \ref{fig:compareLowz}).
The best-fitting models of both low-redshift solutions yield multiple 2--5$\sigma$ deviations from the photometric measurements resulting in much larger $\chi^2$ values (39.2--63.1) relative to the best-fitting $z\simeq6.8$ solution ($\chi^2 = 4.2$; Fig. \ref{fig:compareLowz}).
The high redshift case is clearly preferred with \textsc{beagle}. 

We also investigate how well a low-redshift ($z<4$) solution can self-consistently explain the optical through sub-mm data at the position of COS-87259.
To do so, we fit this full suite of photometry using \textsc{x-cigale} \citep{Boquien2019,Yang2020_xcigale} which directly links the rest-UV/optical dust attenuation to the IR emission via energy balance.
While \textsc{beagle} finely samples the (optical/near-IR) photometric redshift prior with a Monte Carlo Markov Chain algorithm, it does not yet implement energy balance when computing the rest-frame IR emission.
We run \textsc{x-cigale} at two separate fixed redshifts set to the best-fitting values from \textsc{beagle}: $z=0.967$ for the dusty, strong emission line solution and $z=1.52$ for the old stellar population solution which yields a prominent Balmer break at $\approx$9600 \AA{}.
For each fit, we adopt the \citet{BruzualCharlot2003} stellar population synthesis models with a \citet{Chabrier2003} IMF, a delayed SFH with an allowed recent (1--10 Myr) constant SFR burst, and a \citet{Calzetti2000} attenuation law.
We apply the dust emission models of \citet{Draine2014} where the dust is assumed to be illuminated by both a diffuse, low-intensity component and a power-law intensity component from star-forming regions.
Finally, we allow for AGN emission using the \textsc{skirtor} models \citep{Stalevski2012,Stalevski2016} where the dusty torus is treated as a clumpy two-phase medium.
The orientation, extent, density profile, and optical depth of the dusty torus are left as free parameters along with the relative luminosity of the AGN as well as the reddening due to polar dust.
We do not incorporate the X-ray or radio data into the \textsc{x-cigale} fits as the input X-ray flux must be corrected for absorption apriori and the radio module in \textsc{x-cigale} does not allow for a possible contribution from AGN synchrotron emission.

The galaxy+AGN emission models of \textsc{x-cigale} are unable to match the optical through sub-mm data of COS-87259 (best-fitting reduced $\chi^2 > 5$). 
As expected from the \textsc{beagle} results, neither a $z\simeq1$ dusty, strong emission line solution nor a $z\simeq1.5$ old stellar population solution can reproduce the sharp flux discontinuity between ib945 and nb973.
But moreover the \textsc{x-cigale} models demonstrate that the far-infrared and sub-mm data of COS-87259 also disfavor $z\simeq1-1.5$ solutions.
While COS-87529 displays a fairly flat spectral shape (in F$_{\nu}$) between 100--850$\mu$m (see Table \ref{tab:photometry}), the best-fitting $z\simeq1-1.5$ models from \textsc{x-cigale} predict that the flux density declines considerably at $\gtrsim$300--400$\mu$m, yielding $\geq$2.5$\sigma$ offsets from the measured 500$\mu$m and 850$\mu$m photometry.
This behavior in the \textsc{x-cigale} models is consistent with the expectation of a $T\gtrsim20$ K starburst dust temperature at $z\simeq1-1.5$ \citep[e.g.][]{Schreiber2018}, such that the Rayleigh-Jeans tail would appear at $\gtrsim$300--400$\mu$m observed-frame.

Our MMT spectra (\S\ref{sec:MMT}) allow us to further assess the plausibility of the $z\simeq1$ dusty strong emission line galaxy solution, which is the most preferred $z<4$ scenario with both \textsc{beagle} and \textsc{x-cigale}.
In this solution, the observed sharp break between ib945 and nb973 is modeled by strong H$\beta$, [OIII]$\lambda$4959,5007, and H$\alpha$ emission boosting the nb973, \textit{y}, \textit{Y}, and \textit{J} photometry above the continuum.
We estimate the likelihood of not detecting any of these lines in our Binospec and MMIRS observations.
Mock emission lines are inserted into our 1D spectra where the wavelength positions and strengths of these mock lines are taken from the photometric redshift and line flux probability distributions output by \textsc{beagle}.
The S/N of each mock emission line is calculated using the 1D noise spectrum (which accounts for skylines) and by integrating over the FWHM of the mock line which we assume ranges between 50--500 km s$^{-1}$.
We find it likely ($>$98\%) that at least one of the four emission lines would have been detected at $>$5$\sigma$ in our Binospec or MMIRS spectra assuming this low-redshift star-forming galaxy solution, which we again emphasize poorly reproduces the observed photometry.
However, given the implied extremely high optical depth of these low-redshift solutions, we acknowledge that the dust attenuation may not follow simple wavelength-dependent screen models \citep[e.g.][]{Chary2007} as we have assumed in our \textsc{beagle} fits.

Another potential low-redshift solution to the sharp flux discontinuity of COS-87259 is the so-called 3000 \AA{} break seen in a small subset of low-redshift AGN \citep{Meusinger2016}.
These sharp breaks at $\approx$3000 \AA{} rest-frame are caused by very broad overlapping absorption lines in AGN spectra from low-ionization species of magnesium and iron.
AGN showing these breaks belong to the FeLoBAL (iron low-ionization broad absorption line) population, which represent a rare ($\approx$2\%) class of luminous quasars (\citealt{Dai2012}).
Using the FeLoBAL spectral library from \citet{Meusinger2016}, we have determined that absorption troughs from such a source could plausibly reproduce the factor of $>$3 flux discontinuity between ib945 and nb973 seen from COS-87259 if at $z\approx2.4$.
However, it is not necessarily clear that a $z\approx2.4$ FeLoBAL would go undetected in all nine HSC and ACS bands blueward of nb973. 
Such a scenario would seem to require a combination of contiguous overlapping absorption troughs and extremely strong reddening blueward of the break.
This FeLoBAL interpretation also cannot easily explain the extended morphology of the \HST{} F160W detection (Fig. \ref{fig:postageStamps}) as it would suggest that the observed near-IR flux is dominated by a point source.

The final possibility we consider is that COS-87259 is a $z\simeq7$ galaxy being gravitationally lensed by a foreground extremely dusty starburst galaxy.
It is conceivable that such a low-redshift source would go undetected in the optical HSC data (where the dropout is seen) but contribute significantly to the VIRCam and IRAC detections and dominate the mid-IR, far-IR, and radio measurements. Currently there is no evidence of lensing in {\it HST} images (Fig. \ref{fig:postageStamps}), but higher resolution data at redder wavelengths are needed to test the lensing morphology. 
Clearly the chance alignment of a massive, dusty star-forming galaxy at low redshift with a galaxy at $z\simeq 6.8$ (both of which are very rare) is unlikely in the COSMOS area.
Nevertheless a spectroscopic redshift will ultimately be required to establish the true nature of COS-87259 beyond any doubt. 

While we did not detect \Lya{} from COS-87259 with Binospec, these observations remain consistent with an unlensed $z\simeq6.8$ solution.
With our current Binospec data, we can only place a 5$\sigma$ \Lya{} EW upper limit of $<$12.5 \AA{} in the cleanest (i.e. skyline-free) regions of the spectrum.
This best-case upper limit is larger than the typical EW of UV-bright $z\simeq7$ Lyman-break galaxies (11 \AA{}; \citealt{Endsley2021_LyA}).
Furthermore, the extremely red rest-UV slope of COS-87259 ($\beta = -0.59$) suggests a large dust content, which would efficiently destroy \Lya{} photons that resonantly scatter within the ISM of the galaxy. 
Given the high SFR we infer for COS-87259 (\S\ref{sec:self_consistent}), ALMA measurements of far-IR cooling lines (e.g. [CII], CO) are likely to provide a much more tractable means of confirming the redshift of this source. 

\subsection{Physical Interpretation of the Multi-wavelength SED: Implications of the \boldmath{$z\simeq 6.8$} Redshift Solution of COS-87259} \label{sec:self_consistent}

\begin{figure}
\includegraphics{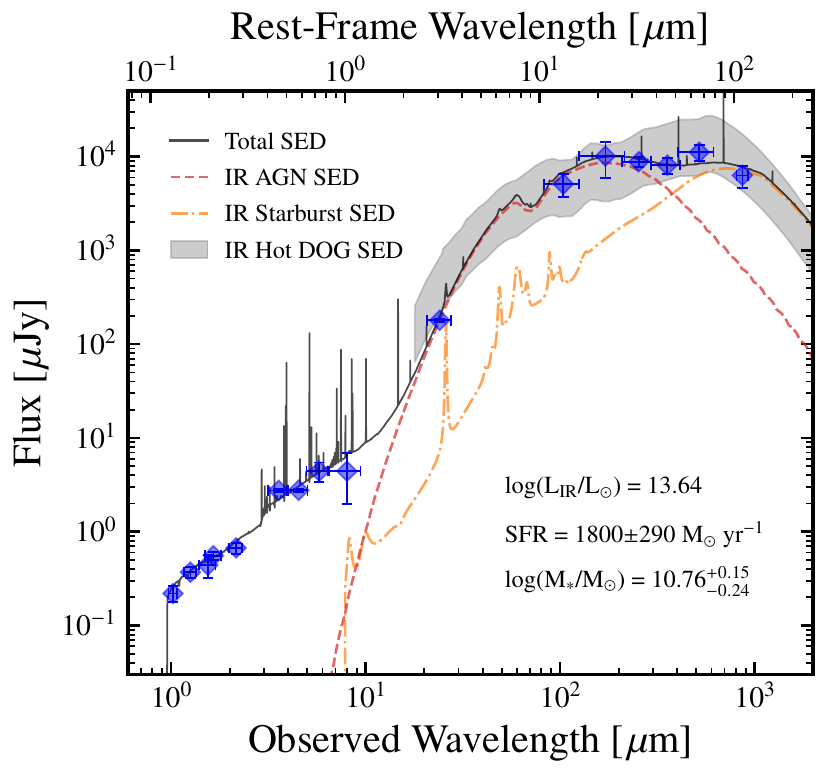}
\caption{Results from fitting the near-IR, mid-IR, and far-IR photometry of COS-87259 with \textsc{x-cigale} \citep{Boquien2019,Yang2020_xcigale}. We find that these data (blue diamonds) can be self-consistently modeled at $z=6.83$ (the photo-z from \textsc{beagle}; see Fig. \ref{fig:BEAGLE}) by a massive galaxy with hyperluminous infrared emission powered by both an intense, highly-obscured starburst as well as an obscured AGN. We show the best-fitting SED in black (reduced $\chi^2$ = 0.45). The infrared continuum SEDs of both the obscured AGN (dashed red) and the obscured starburst (dot-dashed orange) from this best-fitting model are also shown. For comparison, we also plot the median SED of hot dust-obscured galaxies (hot DOGs) from \citet{Fan2016_hotDOG} that has been redshifted to $z=6.83$ and normalized to the SPIRE 250$\mu$m measurement. The gray shaded region shows a $\pm$0.3 dex scatter about this median SED (the approximate dispersion in observed hot DOG SEDs; \citealt{Tsai2015,Fan2016_hotDOG}).}
\label{fig:cigale}
\end{figure}

We now investigate whether the full multi-wavelength (X-ray through radio) data at the position of COS-87259 can be self-consistently explained by a high-redshift ($z\simeq6.8$) solution. 
We fit the photometry of COS-87259 using \textsc{x-cigale} following the same procedure as described in \S\ref{sec:photoz} except that here we fix the redshift to $z=6.83$ (the photometric redshift from \textsc{beagle}).
Because the redshift is fixed, we do not include the HSC and ACS flux density measurements as these bands are impacted by IGM absorption. 
We also do not include the radio photometry as \textsc{x-cigale} assumes this emission arises from star formation using local far-IR to radio correlations (we discuss the likely origin of the radio emission below).

The near-, mid-, and far-infrared emission of COS-87259 can be self-consistently modeled with \textsc{x-cigale} at $z=6.83$ (Fig. \ref{fig:cigale}). 
The extremely red rest-UV slope and sub-mm flux density measurements are simultaneously reproduced by a highly-obscured starburst with a SFR of 1800$\pm$290 \Msol{} yr$^{-1}$ over the past 10 Myr.
Such a large SFR is similar to that inferred for spectroscopically-confirmed sub-mm galaxies at $z>6$ (400--3400 \Msol{} yr$^{-1}$; \citealt{Riechers2013,Marrone2018,Zavala2018}).
The total stellar mass inferred by the \textsc{x-cigale} fit is \logMstar{} = 10.76$^{+0.15}_{-0.24}$, consistent with that inferred from fitting the optical and near-IR photometry with \textsc{beagle} (\logMstar{} = 10.66$^{+0.28}_{-1.52}$).

The \textsc{x-cigale} fit also demonstrates that the MIPS and PACS measurements can be explained by hot ($T\sim300$ K) dust emission from a torus surrounding an AGN.
Due to the steep mid-infrared slope between the MIPS and PACS bands, the AGN is inferred to be highly obscured with an optical depth at rest-frame 9.7$\mu$m of $\tau_{_{\mathrm{9.7\mu m}}} = 7.7\pm2.5$ and an edge-on viewing angle to the torus of $i = 68\pm17$ degrees (where $i=90$ degrees is defined as a sightline through the torus equator).
In the best-fitting \textsc{x-cigale} model (reduced $\chi^2$ = 0.45), the AGN is inferred to contribute 75\% of the total IR luminosity (\logLIR{} = 13.64 integrated from 8--1000$\mu$m rest-frame) with dust emission from star formation contributing the remaining 25\%.
The rest-frame 6$\mu$m luminosity of COS-87259 is also inferred to be L$_{\mathrm{6\mu m}} = 10^{46.9}$ erg s$^{-1}$ from the best-fitting model, suggesting that this source possesses one of the most mid-infrared luminous AGN known if at $z\simeq6.8$ \citep{Stern2015,Martocchia2017}.

While \textsc{x-cigale} allows one to fit SEDs with a large degree of flexibility in parameter space, it is also worth verifying whether the mid+far-IR SED of COS-87259 can be reasonably well fit by a combination of empirical AGN and starburst templates assuming $z=6.83$. 
To this end, we adopt the warm dust-deficient AGN SED from \citet{Lyu2017} along with the starburst SED of Haro 11 from \citet{Lyu2016}, the latter of which has been shown to reproduce the far-IR data of multiple massive $5<z<7$ galaxies \citep{DeRossi2018}.
Using a custom SED fitting code (Lyu et al. in prep), we find that the MIPS, PACS, SPIRE, and SCUBA-2 data of COS-87259 can be reasonably well reproduced by a combination of these two empirical templates with a reduced $\chi^2$ of 1.4.
The inferred properties of COS-87259 (e.g. total infrared luminosity, star formation rate, AGN optical depth) from this fitting approach agree within $\lesssim$0.3 dex with that output by \textsc{x-cigale}.

The hyperluminous infrared emission (\logLIR{}$>$13) and steep mid-infrared slope of COS-87259 suggests that this source may be a higher-redshift analog of $z\sim1-4.5$ hot dust-obscured galaxies (hot DOGs) discovered from \textit{WISE} data (e.g. \citealt{Eisenhardt2012,Wu2012,Tsai2015}).
Similar to COS-87259, these lower-redshift hot DOGs are found to possess very large stellar mass (\logMstar{}$\sim$11--12), heavily obscured AGN, and often recent intense starbursts (SFR$\sim$200-3000 \Msol{} yr$^{-1}$; \citealt{Assef2015,DiazSantos2021}).
The SED shape of COS-87259 at rest-frame $\approx$3--100$\mu$m is also consistent with that of known hot DOGs (Fig. \ref{fig:cigale}), further supporting this comparison.

The X-ray non-detection at the position of COS-87259 is consistent with this highly obscured AGN interpretation.
Lower redshift hot DOGs have been found to exhibit relatively weak intrinsic (i.e. absorption-corrected) X-ray flux at a fixed mid-infrared luminosity compared to type 1 AGN \citep{Ricci2017,Vito2018}.
Assuming a standard AGN power-law photon index of $\Gamma = 1.8$, the \textit{Chandra} non-detection of COS-87259 implies an upper limiting X-ray luminosity of L$_{X} < 10^{44.8}$ erg s$^{-1}$ in the rest-frame 10--40 keV range.
This is similar to upper limiting X-ray luminosities of lower-redshift hot DOGs with comparable mid-infrared luminosities as COS-87259 (L$_{\mathrm{6\mu m}} \approx 10^{47}$ erg s$^{-1}$; \citealt{Stern2014}).
Here, we are focusing our attention on the hard 10--40 keV X-ray regime for two reasons.
The first is that the \textit{Chandra} data closely probes this rest-frame energy range (assuming $z\simeq6.8$) and the second is that such hard X-rays are less impacted by Compton-thick absorption associated with large hydrogen column densities ($N_H \gtrsim 10^{24}$ cm$^{-2}$).
The high optical depth inferred from \textsc{x-cigale} ($\tau_{_{\mathrm{9.7\mu m}}} = 7.7\pm2.5$) suggests that the AGN within COS-87259 is likely Compton-thick \citep{Shi2006} which would yield $\gtrsim$0.25 dex absorption in the rest-frame 10--40 keV luminosity \citep{Stern2014,Lansbury2015}.

The radio properties of COS-87259 further support the presence of an AGN in this system (if at $z\simeq6.8$). 
The observed 1.32 GHz flux density is approximately 45$\times$ higher than that expected for synchrotron emission from star formation \citep{Mancuso2015} assuming SFR$\approx$1800 \Msol{} yr$^{-1}$ (as inferred from \textsc{x-cigale}), suggesting a dominant contribution of non-thermal emission from an AGN.
It is also worth noting that the VLA and MeerKAT flux density measurements may be underestimating the true radio luminosity from COS-87259.
At $z\simeq6.8$, inverse Compton scattering is expected to attenuate high-frequency synchrotron emission due to the hot cosmic microwave background \citep[e.g.][]{Krolik1991}.
The impact of this effect is perhaps evidenced by the ultra-steep slope measured for COS-87259 between the 1.4 and 3 GHz bands ($\alpha = -2.06^{+0.27}_{-0.25}$), which is much steeper than that of typical sources in the VLA-COSMOS survey ($\alpha = -0.7$; \citealt{Smolcic2017}; see also \citealt{An2021}).
Accounting for this inverse Compton scattering effect would only decrease the likelihood that the radio emission of COS-87259 is due to star formation.
Moreover, if the starburst in COS-87259 is very young ($\lesssim$5 Myr), there may be a significant fraction of recently formed high-mass ($>$8 \Msol{}) stars that have yet to produce supernovae, lowering the expected synchrotron emission from star formation activity in this system \citep[e.g.][]{Mancuso2015}.
We have assumed that the VLA and MeerKAT detections are dominated by synchrotron emission given that the ultra-steep slope measured at $\geq$1.3 GHz ($-2.1 \lesssim \alpha \lesssim -1.6$) is inconsistent with that expected from thermal free-free emission at these high frequencies ($\alpha \sim -0.1$; e.g. \citealt{Rubin1968}).

To further assess the plausibility that COS-87259 may host a radio AGN at $z\simeq6.8$, we compare the radio properties of COS-87259 with those of known high-redshift ($z=4-6$) radio galaxies.
The observed extent of the VLA 3 GHz detection (1.04\arcsec{}; \S\ref{sec:radio}) translates to a projected physical size of 5.5 kpc assuming $z=6.83$, consistent with measurements from the two spectroscopically-confirmed $z=5-6$ radio galaxies with high-resolution imaging (3.5--7.4 kpc; \citealt{vanBreugel1999,Saxena2018}).
The radio emission coincident with COS-87259 also shows a spectral steepening towards higher frequencies (Fig. \ref{fig:radioSlope}), consistent with the behavior of many spectroscopically-confirmed $z>4$ radio galaxies \citep[e.g.][]{Ker2012,Saxena2018_faintRadio,Saxena2018,Yamashita2020}.
For example, GLEAM 0856 at $z=5.55$ exhibits an ultra-steep ($\alpha = -1.51$) radio slope at $\gtrsim$1.4 GHz yet a relatively flat slope ($\alpha = -0.78$) at $\lesssim$1 GHz (Fig. \ref{fig:radioSlope}; \citealt{Drouart2020}).
While the observed 1.4 GHz flux density of COS-87259 is $\approx$1000$\times$ fainter than that of most currently known $z>4$ radio galaxies \citep{Saxena2018_faintRadio,Saxena2018,Saxena2019,Drouart2020,Yamashita2020}, this can be explained by the fact that $z>4$ radio galaxies are often identified from all-sky surveys that probe much higher radio luminosities relative to the extremely deep VLA-COSMOS data.
There are many radio galaxies at lower redshifts ($z<4$) with rest-frame 1.4 GHz luminosities comparable to that of COS-87259 (log[L$_{\mathrm{1.4\ GHz}}/(\mathrm{W\ Hz}^{-1})$] = $25.4\pm0.2$; e.g. \citealt{Best2012,Simpson2012,Rigby2015}).
Here, we have adopted the low-frequency radio slope of $\alpha = -0.86^{+0.22}_{-0.16}$ at observed frequencies lower than 1.32 GHz to compute L$_{\mathrm{1.4\ GHz}}$ for COS-87259 (see Fig. \ref{fig:radioSlope}).
Finally, we note that the high stellar mass of COS-87259 (\logMstar{} = 10.8) is consistent with the massive (\logMstar{} $\sim$ 11.0-11.5) nature of known radio galaxies at $z\lesssim5$ \citep[e.g.][]{Saxena2019}, where the stellar mass is found to correlate weakly with radio luminosity at L$_{\mathrm{1.4\ GHz}}$ $\gtrsim$ 10$^{24}$ W Hz$^{-1}$ \citep[e.g.][]{Seymour2007,Best2012}.

The rest-frame 1.4 GHz luminosity of COS-87259 (L$_{\mathrm{1.4\ GHz}}$ $\approx$ 10$^{25.4}$ W Hz$^{-1}$) suggests that this source falls under the class of radio-loud AGN (L$_{\mathrm{1.4\ GHz}}$ $>$ 10$^{25}$ W Hz$^{-1}$; e.g. \citealt{Fanaroff1974}).
Another approach to classifying the radio-loudness of high-redshift AGN is to compare their radio flux density to that in the rest-UV, where a source is considered radio loud if $R_{2500} \equiv F_{\mathrm{5\, GHz}}/F_{\mathrm{2500\,\mathring{A}}} > 10$ \citep[e.g.][]{Kellermann1989}. 
For COS-87259, we calculate $R_{2500} \approx 180$ where we use a power-law fit to the VIRCam data to derive $F_{\mathrm{2500\,\mathring{A}}}$ = 0.65 $\mu$Jy and estimate $F_{\mathrm{5\, GHz}} \approx 115$ $\mu$Jy assuming $\alpha \approx -0.86$ at observed frequencies lower than 1.32 GHz, consistent with the MeerKAT, LOFAR, and GMRT data (see Fig. \ref{fig:radioSlope}).
This $R_{2500} \approx 180$ value also suggests that the AGN within COS-87259 is radio loud.

To place COS-87259 into context of known type 1 AGN at $z\sim7$ (i.e. quasars), we estimate the extinction-corrected UV magnitude of its AGN.
From the \textsc{x-cigale} fit, the intrinsic bolometric luminosity of the AGN accretion disk is inferred to be log$\left(\mathrm{L_{AGN}}/\mathrm{L_{\odot}}\right) = 13.6\pm0.1$.
We convert this into a UV luminosity (at rest-frame 1450 \AA{}) using the relation from \citet{Runnoe2012}, finding that COS-87259 would appear as an extremely UV-luminous quasar with $\Muv{} \approx -27$ if unobscured (c.f. the observed $\Muv{} = -21.7$).
Such UV-luminous $z\sim7$ quasars are often found to be powered by supermassive black holes with M$_{\mathrm{BH}} \gtrsim 10^9 \Msol{}$ \citep[e.g.][]{Mortlock2011,Banados2018,Yang2020,Wang2021}.
The very large inferred bolometric AGN luminosity of COS-87259 also suggests that a 1.2$\times$10$^9$ \Msol{} black hole is present within this system (if at $z\simeq6.8$) assuming Eddington-limited accretion (i.e. $\lambda_E = 1$).
While this estimated black hole mass is substantially higher than what would be inferred using the local M$_{\mathrm{BH}}$--\Mstar{} relation (M$_{\mathrm{BH}} \sim 10^8 \Msol{}$; \citealt{HaringRix2004}), we note that many high-redshift quasars have been found to harbor black holes much more massive than that expected from their host galaxy mass \citep[e.g.][]{Trakhtenbrot2015,Venemans2016,Pensabene2020}.
Lower-redshift hot DOGs also exhibit enhanced AGN luminosities relative to that expected from their stellar masses \citep{Assef2015,Tsai2015}, suggesting that these highly-obscured AGN are either accreting significantly above the Eddington limit (see also \citealt{Ferris2021}) and/or lie above the local M$_{\mathrm{BH}}$-\Mstar{} relation.

Having now placed COS-87259 into context of the UV-luminous $z\sim7$ quasar population, we can begin to assess the likelihood of identifying this source within the 1.5 deg$^2$ COSMOS field.
The volume probed by COSMOS between $z=6.6-6.9$ (the approximate redshift selection interval of \citetalias{Endsley2021_OIII}; see \S\ref{sec:sample}) is 0.0036 comoving Gpc$^{3}$ while extremely UV-luminous ($\Muv{} < -27$) quasars at $z\simeq6.75$ have a space density of 0.77 per comoving Gpc$^{3}$ \citep{Jiang2016}.
This indicates that the odds of identifying an $\Muv{} < -27$ quasar in the COSMOS field between $z=6.6-6.9$ is very unlikely ($\sim$0.03\%), suggesting that COS-87259 was an extremely lucky find.
However, we note that this likelihood increases with the assumed fraction of highly obscured quasars in the very early Universe.
Recent studies have shown that a subset of UV-luminous $z>6$ quasars exhibit much smaller \Lya{} proximity zones than expected based on their inferred black hole mass, consistent with a picture wherein their black holes grew during highly obscured phases for a large fraction ($\gtrsim$95\%) of their lifetime \citep{Davies2019,Eilers2020,Eilers2021}.
High-resolution hydrodynamic simulations also predict that supermassive black holes within massive $z>7$ galaxies were perhaps very often ($\sim$99\%) obscured by the dense gas content of their host galaxies \citep{Trebitsch2019,Ni2020}.
However, such large obscured fractions at $z\sim7$ would necessitate a very rapid change in the gas properties of galaxies from $z\sim6$ where the obscured fraction is found to be $\sim$70--80\% based on X-ray studies \citep{Vito2018_obscuredFraction}.
Larger-area multi-wavelength surveys will ultimately be required to determine the fraction of obscured extremely luminous quasars at $z\sim7$.

If $\sim$billion solar mass black holes at $z\sim7$ are indeed obscured for 95--99\% of their lifetime, the odds of identifying COS-87259 within the COSMOS field would increase substantially, though still only to $\sim$1\%.
However, the presence of a strong $z\simeq6.6-6.9$ photometric overdensity around COS-87259 (see \S\ref{sec:overdensity}) suggests that the region containing this source is special and may be preferentially traced by an extremely massive halo.
Nonetheless, we again emphasize that a spectroscopic redshift will be necessary to determine whether COS-87259 in fact lies at $z\simeq6.8$ as suggested by the optical/near-IR data.

The parent sample of 41 UV bright ($\Muv{} \lesssim -21.25$) galaxies from which COS-87259 was selected \citepalias{Endsley2021_OIII} also provides some insight into how this system compares to the general population of UV-luminous $z\simeq 7$ galaxies. 
COS-87259 is the only source in this sample with any \Spitzer{}/MIPS, \textit{Herschel}, JCMT, or VLA detection. 
COS-87259 is also the reddest object ($\beta=-0.6$) and one of only three galaxies in the sample known to have stellar mass above 10$^{10}$ \Msol{}. 
The exact number of galaxies in the parent sample above this mass threshold is uncertain as 11 of 41 galaxies in \citetalias{Endsley2021_OIII} have significantly confused IRAC photometry.
Assuming these 11 sources have a similar stellar mass distribution as the 30 with robust IRAC photometry, we infer that 4 of 41 galaxies in  \citetalias{Endsley2021_OIII} have stellar mass above 10$^{10}$ \Msol{}.
If the redshift of COS-87259 is confirmed at $z\simeq 6.8$, it would indicate that at least 1 of these 4 very massive UV-luminous systems exhibit both AGN and highly-obscured star formation activity.

As discussed above, it is likely that COS-87259 is an exceptional system if at $z\simeq6.8$.
It is nonetheless worth noting that the strong sub-mm flux density of COS-87259 is consistent with a scenario in which highly-obscured star formation activity is more common among the most massive ($>$10$^{10}$ \Msol{}) early galaxies, which may potentially influence current estimates of the star formation rate budget at $z>6$ (see e.g. \citealt{Zavala2021}). 
To provide a qualitative estimate of this potential impact, we note that the star formation rate of COS-87259 alone (SFR$\approx$1800 \Msol{} yr$^{-1}$) is equivalent to $\approx$3--4\% of the total SFR from all $\Muv{} < -17$ $z=6.6-6.9$ galaxies in the 1.5 deg$^2$ COSMOS field (i.e. those which would be detectable with ultra-deep \HST{} imaging).
Here, we have assumed the UV-based $z\sim7$ cosmic star formation rate density results of \citet{Bouwens2015_LF} and \citet{Finkelstein2015_LF}.
COS-87259 may also suggest that AGN activity is somewhat common among the most massive, UV-luminous galaxies in the reionization era.
Several other UV-bright galaxies at $z\sim7-9$ have shown evidence of significant AGN activity via detections of NV$\lambda$1240 emission \citep{Tilvi2016,Laporte2017,Mainali2018,Endsley2021_LyA}, perhaps indicating that AGN make a non-negligible contribution to the ionizing photon budget at early times \citep{Madau2015}.
Notably, radio AGN appear to be very rare among narrow-band selected Ly$\alpha$ emitters at $z\simeq5.7$ and $z\simeq6.6$ \citep{Gloudemans2021}, consistent with a scenario in which such bright Ly$\alpha$ emitters more often trace young unobscured starbursts rather than the most massive galaxies harboring luminous AGN.
Wider-area mutli-wavelength coverage will be required to better determine the general properties of massive reionization-era galaxies, including those which are too reddened to be selected in the rest-UV.

\subsection{Photometric \boldmath{$z\simeq6.6-6.9$} Overdensity Around COS-87259} \label{sec:overdensity}

Both radio galaxies and hot dust-obscured galaxies are often found to reside in highly overdense environments \citep[e.g.][]{Pentericci2000,Miley2004,Venemans2007,Jones2014,Jones2015,Fan2017_hotDOG}.
We investigate whether COS-87259 may reside in an overdense environment by comparing the surface density of $z\simeq6.6-6.9$ Lyman-break galaxy candidates around this source to the average surface density across COSMOS. 
The $z\simeq6.6-6.9$ Lyman-break galaxy sample used in this analysis are identified using the selection criteria of \citetalias{Endsley2021_OIII}.
To provide the best statistics of neighboring galaxies, we ignore the magnitude cuts from \citetalias{Endsley2021_OIII} (i.e. \textit{J}$<$25.7 or \textit{K}$_s\!\!<$25.5) allowing for fainter ($-21 \lesssim \Muv{} \lesssim -20.5$) sources to enter the sample.
This selection results in a total of 67 $z\simeq6.6-6.9$ galaxy candidates across the ultra-deep UltraVISTA stripes (0.73 deg$^2$ total\footnote{\url{http://ultravista.org/}}) corresponding to an average surface density of 0.0255 sources per arcmin$^2$.
We consider only the area of the ultra-deep stripes because COS-87259 resides within one of these stripes, and the surface density of $z\simeq6.6-6.9$ candidates is much lower in the UltraVISTA deep stripes which are $\sim$1 mag shallower in the VIRCam data.

In an 18 arcmin$^2$ region around COS-87259, we identify five $J<25.8$ ($\Muv{} < -21$) $z\simeq6.6-6.9$ galaxy candidates (including COS-87259) yielding a local surface density of 0.278 arcmin$^{-2}$.
This is approximately 11$\times$ the average surface density of $z\simeq6.6-6.9$ candidates across the ultra-deep UltraVISTA stripes, suggesting that COS-87259 may trace a highly overdense region at $z\simeq6.8$. 
For comparison, radio and hot dust-obscured galaxies at $z>2$ are typically found to reside in environments with $\sim3-5\times$ the number of neighboring sources relative to blank fields \citep[e.g.][]{Venemans2007,Jones2014,Fan2017_hotDOG}.
Such strong overdensities are, at $z\gtrsim7$, likely to carve out large ionized regions in the mostly neutral IGM. 
Further spectroscopy (targeting the fainter photometric neighbors) will be able to verify whether COS-87259 is tracing such an overdense ionized region \citep[e.g.][]{Endsley2022_bubble}.

This strong overdensity suggests that COS-87259 (if at $z\simeq6.8$) likely occupies a special region of the early Universe that may be preferentially traced by an extremely massive halo (and hence an exceptionally luminous AGN).
To test this, we analyze the environments of the most massive halos within the very large (1 Gpc/$h$)$^3$ MultiDark Planck 2 simulation (MDPL2; \citealt{Klypin2016}) at a snapshot redshift of $z=6.85$.
We generate multiple lines of sight through the 10 most massive halos in the simulation and ask in how many realizations there are at least four neighboring $z=6.6-6.9$ sources with $\Muv{} < -21$ within an 18 arcmin$^2$ region (where the central massive halo is assumed to lie at $z=6.75$).
Here, halos are assigned UV magnitudes using the \textsc{universemachine} model \citep{Behroozi2019} which has been calibrated to empirical UV luminosity functions, \Muv{}-\Mstar{} relations, sSFR, and cosmic star formation rate density measurements at high redshifts ($z\sim4-10$). 
We find that in 66\% of realizations, the ten most massive halos in MDPL2 occupy overdensities similar to that of COS-87259.
This is a marked contrast to only 0.10\% of randomly positioned and oriented sightlines that contain five $\Muv{} < -21$ $z=6.6-6.9$ galaxies in an 18 arcmin$^2$ region.
This suggests that extremely massive halos in the early Universe would indeed preferentially trace the strong overdensity seen around COS-87259.

\section{Summary and Outlook} \label{sec:summary}

We have reported mid-IR, far-IR, sub-mm, and radio detections coincident with the position of a UV-luminous ($\Muv{} = -21.7$) $z\simeq7$ Lyman-break galaxy candidate located in the 1.5 deg$^2$ COSMOS field.
This source, COS-87259, exhibits a sharp flux discontinuity (factor $>$3) between the HSC ib945 and nb973 bands ($\Delta \lambda = 250$ \AA{}) and is undetected in all nine bands blueward of 0.96$\mu$m (including \HST{} ACS/F814W) as expected from a Ly$\alpha$ break at $z\simeq6.8$. 
The full multi-wavelength data (X-ray through radio) of COS-87259 can be self-consistently explained by a massive (\Mstar{} = 10$^{10.8}$ \Msol{}) and extremely red ($\beta = -0.59$) galaxy at $z=6.83\pm0.06$ with hyperluminous infrared emission (\logLIR{} = 13.6) powered by both an intense burst of highly-obscured star formation (SFR $\approx$ 1800 \Msol{} yr$^{-1}$) and an obscured ($\tau_{_{\mathrm{9.7\mu m}}} = 7.7\pm2.5$) radio-loud (L$_{\mathrm{1.4\ GHz}}\approx10^{25.4}$ W Hz$^{-1}$) AGN.
The radio emission is compact (1.04$\pm$0.12 arcsec) and exhibits an ultra-steep spectrum between 1.32--3 GHz ($\alpha=-1.57^{+0.22}_{-0.21}$) that flattens at lower frequencies ($\alpha = -0.86^{+0.22}_{-0.16}$ between 0.144--1.32 GHz), consistent with known $z>4$ radio galaxies.
We also find evidence that COS-87259 may reside in significantly overdense (11$\times$) environment at $z\simeq6.8$, as is common for systems hosting radio-loud AGN at lower redshifts.

We consider models of low-redshift galaxies and find that such solutions are unlikely to reproduce the optical and near-infrared data, at least when assuming a suite of standard dust screen models.
It is possible that the sharp flux discontinuity seen in COS-87259 can be produced by a very rare class of AGN (FeLoBAL) if at $z\simeq2.4$.
However, it is not immediately clear that such an AGN would go undetected in all nine bands blueward of 0.96$\mu$m.
This low-redshift FeLoBAL solution also cannot easily explain the extended morphology of the \HST{} WFC3/F160W detection.
We also cannot rule out the possibility that COS-87259 is a $z\simeq6.8$ source that is being gravitationally lensed by a foreground extremely dusty starburst galaxy, the latter of which could dominate the mid-IR through radio emission.
A spectroscopic redshift will ultimately be required to establish the true nature of COS-87259 beyond any doubt.
While optical (0.75--1.00$\mu$m) spectroscopic observations of COS-87259 has resulted in no detection of \Lya{} emission, this is consistent with expectations of a $z\simeq6.8$ solution given the implied extremely red rest-UV slope of this source.
Observations of far-IR cooling lines (e.g. [CII]) would likely offer a better means of confirming the redshift of COS-87259 due to its high star formation rate.

If COS-87259 is confirmed to lie at $z\simeq6.8$, we estimate that at least one of four very massive ($\Mstar{} > 10^{10} \Msol{}$) UV-bright $z\simeq6.6-6.9$ galaxies in COSMOS would be known to exhibit both AGN and highly-obscured star formation activity.
Such a confirmation would be consistent with a picture wherein very massive galaxies contribute significantly to the cosmic star formation rate density at $z>6$.
The inferred star formation rate of COS-87259 alone (SFR$\approx$1800 \Msol{} yr$^{-1}$) is equivalent to $\approx$3--4\% of the total SFR from all $\Muv{} < -17$ $z\simeq6.6-6.9$ galaxies in the 1.5 deg$^2$ COSMOS field (i.e. those detectable with ultra-deep \HST{} imaging; \citealt{McLure2013,Bouwens2015_LF,Finkelstein2015_LF}).
If confirmed, COS-87259 may also suggest that AGN are fairly common among the most massive, UV-luminous galaxies in the reionization era.
Several other UV-bright galaxies at $z\sim7-9$ have shown evidence of significant AGN activity via detections of NV$\lambda$1240 emission \citep{Tilvi2016,Laporte2017,Mainali2018,Endsley2021_LyA}, perhaps indicating that AGN make a non-negligible contribution to the ionizing photon budget at early times \citep{Madau2015}.
Wider-area multi-wavelength surveys will ultimately be required to better characterize the demographics of AGN as well as the incidence of intense star formation activity within very massive reionization-era galaxies, including those which are too reddened to be selected in the rest-frame ultraviolet.

\section*{Acknowledgements}

The authors sincerely thank the anonymous referee for their helpful and constructive comments.
We also kindly thank the LoTSS collaboration (and especially Ken Duncan) for allowing us to use the LOFAR 144 MHz data on the source that is the dominant subject of this work. 
RE and DPS acknowledge funding from NASA JWST/near-IRCam contract to the University of Arizona, NAS5-02015. DPS acknowledges support from the National Science Foundation through the grant AST-2109066. XF acknowledges support from the US NSF Grant AST 19-08284. RS acknowledges support from a STFC Ernest Rutherford Fellowship (ST/S004831/1). FW acknowledges the support provided by NASA through the NASA Hubble Fellowship grant \#HST-HF2-51448.001-A awarded by the Space Telescope Science Institute, which is operated by the Association of Universities for Research in Astronomy, Incorporated, under NASA contract NAS5-26555. RJB acknowledges support from TOP grant TOP1.16.057 provided by the Nederlandse Organisatie voor Wetenschappelijk Onderzoek (NWO).

LOFAR data products were provided by the LOFAR Surveys Key Science project (LSKSP; \url{https://lofar-surveys.org/}) and were derived from observations with the International LOFAR Telescope (ILT). LOFAR \citep{vanHaarlem2013} is the Low Frequency Array designed and constructed by ASTRON. It has observing, data processing, and data storage facilities in several countries, which are owned by various parties (each with their own funding sources), and which are collectively operated by the ILT foundation under a joint scientific policy. The efforts of the LSKSP have benefited from funding from the European Research Council, NOVA, NWO, CNRS-INSU, the SURF Co-operative, the UK Science and Technology Funding Council and the J\"{u}lich Supercomputing Centre.

This work is based [in part] on observations made with the Spitzer Space Telescope, which was operated by the Jet Propulsion Laboratory, California Institute of Technology under a contract with NASA. 
Herschel is an ESA space observatory with science instruments provided by European-led Principal Investigator consortia and with important participation from NASA.
Based on data products from observations made with ESO Telescopes at the La Silla Paranal Observatory under ESO programme ID 179.A-2005 and on data products produced by CALET and the Cambridge Astronomy Survey Unit on behalf of the UltraVISTA consortium.
This research has made use of data obtained from the Chandra Source Catalog, provided by the Chandra X-ray Center (CXC) as part of the Chandra Data Archive.
This research has made use of the NASA/IPAC Infrared Science Archive, which is funded by the National Aeronautics and Space Administration and operated by the California Institute of Technology.

Observations reported here were obtained at the MMT Observatory, a joint facility of the University of Arizona and the Smithsonian Institution.
This paper uses data products produced by the OIR Telescope Data Center, supported by the Smithsonian Astrophysical Observatory.
Based on data products from observations made with ESO Telescopes at the La Silla Paranal Observatory under ESO programme ID 179.A-2005 and on data products produced by CALET and the Cambridge Astronomy Survey Unit on behalf of the UltraVISTA consortium.
RE sincerely thanks the MMT queue observers, Ryan Howie and ShiAnne Kattner, for their assistance in collecting the Binospec and MMIRS data, as well as Ben Weiner and Joannah Hinz for managing the queues.
RE also thanks Mengtao Tang for their guidance in planning the MMIRS observations, as well as Sean Moran for their assistance in the MMIRS mask design.

The Hyper Suprime-Cam (HSC) collaboration includes the astronomical communities of Japan and Taiwan, and Princeton University. The HSC instrumentation and software were developed by the National Astronomical Observatory of Japan (NAOJ), the Kavli Institute for the Physics and Mathematics of the Universe (Kavli IPMU), the University of Tokyo, the High Energy Accelerator Research Organization (KEK), the Academia Sinica Institute for Astronomy and Astrophysics in Taiwan (ASIAA), and Princeton University. Funding was contributed by the FIRST program from the Japanese Cabinet Office, the Ministry of Education, Culture, Sports, Science and Technology (MEXT), the Japan Society for the Promotion of Science (JSPS), Japan Science and Technology Agency (JST), the Toray Science Foundation, NAOJ, Kavli IPMU, KEK, ASIAA, and Princeton University. 
This paper makes use of software developed for Vera C. Rubin Observatory. We thank the Rubin Observatory for making their code available as free software at http://pipelines.lsst.io/.
This paper is based on data collected at the Subaru Telescope and retrieved from the HSC data archive system, which is operated by the Subaru Telescope and Astronomy Data Center (ADC) at NAOJ. Data analysis was in part carried out with the cooperation of Center for Computational Astrophysics (CfCA), NAOJ. We are honored and grateful for the opportunity of observing the Universe from Maunakea, which has the cultural, historical and natural significance in Hawaii. 

This research made use of \textsc{astropy}, a community-developed core \textsc{python} package for Astronomy \citep{astropy:2013, astropy:2018}; \textsc{matplotlib} \citep{Hunter2007_matplotlib}; \textsc{numpy} \citep{harris2020_numpy}; and \textsc{scipy} \citep{Virtanen2020_SciPy}.

\section*{Data Availability}

Many of the near-IR, mid-IR, far-IR and radio catalogs and data underlying this article are available through the NASA/IPAC Infrared Science Archive at \url{https://irsa.ipac.caltech.edu/data/COSMOS/}.
In addition, the HELP and SCUBA-2 catalogs are available at \url{http://hedam.lam.fr/HELP/} and \url{https://vizier.u-strasbg.fr/viz-bin/VizieR?-source=J/ApJ/880/43}, respectively.
The optical HSC data can be found at \url{https://hsc-release.mtk.nao.ac.jp/doc/} and the WFC3/F160W data can be found via the COSMOS-DASH webpage at \url{https://archive.stsci.edu/hlsp/cosmos-dash}.
The \textit{Chandra} Source Catalog v2.0 is available at \url{https://cxc.cfa.harvard.edu/csc/}.
The MeerKAT and LOFAR data are available as part of the MIGHTEE Early Release \citep{Heywood2022} and LoTSS DR2 \citep{Shimwell2022}, respectively.
The MMT spectra will be shared upon reasonable request to the corresponding author.

%%%%%%%%%%%%%%%%%%%%%%%%%%%%%%%%%%%%%%%%%%%%%%%%%%

%%%%%%%%%%%%%%%%%%%% REFERENCES %%%%%%%%%%%%%%%%%%

% The best way to enter references is to use BibTeX:

\bibliographystyle{mnras}
\bibliography{paper_ref} % if your bibtex file is called example.bib

%%%%%%%%%%%%%%%%%%%%%%%%%%%%%%%%%%%%%%%%%%%%%%%%%%

%%%%%%%%%%%%%%%%% APPENDICES %%%%%%%%%%%%%%%%%%%%%

\appendix
 
%%%%%%%%%%%%%%%%%%%%%%%%%%%%%%%%%%%%%%%%%%%%%%%%%%

% Don't change these lines
\bsp	% typesetting comment
\label{lastpage}
\end{document}